\documentclass[useAMS,fleqn,usenatbib]{mnras}
\usepackage{multirow}
\usepackage[T1]{fontenc}
\usepackage{ae,aecompl}
\usepackage{graphicx}
\usepackage{hyperref}
\usepackage{amsmath}	% Advanced maths commands
\usepackage{amssymb}	% Extra maths symbols
\usepackage{threeparttable}
\usepackage{booktabs}

\usepackage{newtxtext,newtxmath}
\usepackage{color}
\usepackage{ulem}

\def\apj {ApJ}
\def\apjl {ApJL}
\def\apjs {ApJS}
\def\aj {AJ}
\def\aap {A\&A}
\def\mnras {MNRAS}
\def\araa {ARA\&A}
\def\nat {Nature}

\def\Msol {h^{-1}{\rm M_{\odot}}}

\def\R200 {R_{200}}

\def\mpc{h^{-1} {\rm{Mpc}}}
\def\kms {\rm{km~s^{-1}}}
\def\Mpc {\rm Mpc}

\def\sag {\textsc{sag}}
\def\mdpl {\textsc{mdpl2}}

\definecolor{vale}{RGB}{147,112,219}
\definecolor{polaco}{RGB}{70,130,80}

\title[Anisotropic infall in the outskirst of clusters]{
Star formation quenching in the infall region around  galaxy clusters
}

\author[Salerno et al.]{\parbox[t]{\textwidth}{Juan Manuel Salerno$^{1}$, Hern\'an Muriel$^{1,2}$, Valeria Coenda$^{1,2}$, Sofía A. Cora$^{3,4}$, Luis Pereyra$^{1}$, Andr\'es N. Ruiz$^{1,2}$, Cristian A. Vega-Martínez$^{5,6}$}
\\
\\
$^{1}$Instituto de Astronom\'ia Te\'orica y Experimental (CCT C\'ordoba, CONICET, UNC), 
Laprida 854, X5000BGR, C\'ordoba, Argentina\\
$^{2}$Observatorio Astron\'omico, Universidad Nacional de C\'ordoba, Laprida 854, X5000BGR, C\'ordoba, Argentina\\
$^{3}$Instituto de Astrof\'isica de La Plata (CCT La Plata, CONICET, UNLP), 
Observatorio Astron\'omico, Paseo del Bosque S/N, B1900FWA,\\
La Plata, Argentina\\
$^{4}$Facultad de Ciencias Astron\'omicas y Geof\'isicas, Universidad Nacional de La Plata, 
Observatorio Astron\'omico,\\
Paseo del Bosque S/N, B1900FWA, La Plata, Argentina, \\
$^{5}$Instituto de Investigaci\'on Multidisciplinar en Ciencia y Tecnolog\'ia, Universidad de La Serena, Ra\'ul Bitr\'an 1305, La Serena, Chile\\
$^{6}$Departamento de Astronom\'ia, Universidad de La Serena, Av. Juan Cisternas 1200 Norte, La Serena, Chile
}

\date{Accepted 2022 October 13. Received 2022 October 13; in original form 2022 May 13}

\pubyear{2022}

\hypersetup{draft} 
\begin{document}
\label{firstpage}
\pagerange{\pageref{firstpage}--\pageref{lastpage}}
\maketitle
 
\begin{abstract}
We analyse the connection between the 
star formation
quenching of galaxies and 
their location in the
outskirts
of clusters in the redshift range $z=[0,2]$ by estimating the fraction of red galaxies. More specifically, we focus on 
galaxies that infall isotropically from those that are infalling alongside filaments. We use a sample of galaxies obtained from the semi-analytic model of galaxy formation \sag~applied to the MultiDark simulation. 
\mdpl.
In agreement with observational results, we find that the infall regions show levels of star formation that are intermediate between those of galaxies in clusters and in the field. Moreover, we show that, in the redshift range [0-0.85], the quenching of the star formation is stronger in the filamentary region than in the isotropic infall region. We also study the fraction of red galaxies as a function of the normalised distance to the cluster centre and find that, for radii $R/R_{\textup{200}}> 3 $, the fraction of red galaxies in the filamentary region is considerably larger than in the isotropic infall region. 
From the analysis of properties of
the main progenitors of galaxies identified
at $z = 0$, we find that they have different evolutionary behaviours depending on the stellar mass and environment. Our results confirm the observational findings that suggest that the infall regions of clusters play an important role in the pre-processing of galaxies along most of the evolutionary history of galaxies. 
\end{abstract}

\begin{keywords}
galaxies: evolution -- galaxies: clusters: general --  
galaxies: star formation -- galaxies: statistics.
\end{keywords}

\section{Introduction}
\label{sect:Intro}

The distribution of galaxies on scales of tens of Mpc presents large anisotropies. The pattern of these anisotropies is known as the cosmic web, and presents structures such as clusters, walls, filaments, and  voids \citep{Bond96, aragon10,Cautun13}. Galaxy clusters are the most massive systems in the Universe, with stellar masses of $\sim  10^{14-15} \,\Msol$ and  velocity dispersions of  $\sim 700-800 \, \kms$. These systems, which are connected between them by filaments, can be in virial equilibrium and are characterised by a deep gravitational potential well and an intracluster medium (ICM) filled with hot ionized gas \citep{Binney:87,Kaiser:86, Dressler:83}.

Dense environments, like galaxy clusters, are the place of several quenching processes that shut down the star formation (SF) activity, and galaxies in these environments tend to be red, early-type, with little or no ongoing SF. These quenching processes can depend either on the stellar mass or on the density of the environment in which galaxies reside, and are referred to as mass and environmental quenching, respectively (\citealt{dressler80,gomez03,martinez08, Peng10}). The latter involve processes such as ram pressure stripping \citep[RPS,][]{GG:1972,Abadi:1999}, tidal stripping \citep[TS,][]{Merritt83} and galaxy harassment \citep{Moore:1999}. These phenomena can remove an important fraction of the hot and cold gas of the cluster galaxies, with the consequent decrease of the star formation rate (SFR).
The removal of the warm gas from the galactic halo by the ICM is known as starvation (e.g.,
\citealt{Larson:1980, Balogh:2000,McCarthy:2008,Bekki:2009,Bahe:2013,Vijayaraghavan:2015}). This process cuts off the supply of gas, thus inhibiting any subsequent SF, a process that can happen even with a gradual removal of the hot gas \citep{cora19}. Even though this mechanism is more effective at the central regions of clusters, it also has been reported in less massive systems (e.g. \citealt{Rasmussen06, Jaffe12}). On the other hand, mass quenching involves internal, self-regulated physical processes, being the most important ones the stellar feedback (e.g., \citealt{Stringer:2012, Bower:2012, Hopkins14, Chan18}) and the feedback from active galactic nuclei (AGN, \citealt{Bower:2012, Hasinger:2008, Silverman:2008}). Stellar feedback is efficient in galaxies with low and intermediate stellar masses \citep{Trujillo-Gomez15}, while at the massive end, the AGN phenomenon is expected to be the dominant effect \citep{Beckmann17}. 

Galaxy quenching has been studied in the state-of-art hydrodynamical cosmological simulations like \textsc{EAGLE}  \footnote{https://eagle.strw.leidenuniv.nl/} \citep{eagle_1,eagle_2} by \citet{pallero_2019} or  \textsc{IllustrisTNG}\footnote{https://www.tng-project.org/} \citep{illustristng} by \citet{donnari_2021} and \citet{walters_2022}, and in cluster resimulations such as \textsc{The Hundred Project}\footnote{https://www.nottingham.ac.uk/~ppzfrp/The300/index.php} \citep{threehundred} by \citet{wang_2018}. This topic has also been addressed by several semi-analytic models of galaxy formation \citep{Henriques17, Stevens17, Cora18, cora19, Xie20}. All of these works stand out the importance of environment in 
galaxy 
quenching and the role of the pre-processing, mostly for satellites or galaxies residing in low mass hosts. On the other hand, SF of central galaxies tend to be quenched in-situ via mass quenching mechanisms like AGN feedback.

Galaxy clusters grow by the accretion of galaxies and groups of galaxies from their outskirts, a phenomenon that occurs preferentially alongside filaments (e.g. \citealt{Colberg:1999, Ebeling:2004, Knebe:2004, Rost:2019, Kuchner:2022}) and, to a lesser extent, from other directions (isotropic infall region). \cite{Martinez16} (hereafter, M16) studied the properties of galaxies in filaments
between groups of galaxies in the Sloan Digital Sky Survey Data Release 7 \citep[SDSS DR7, ][]{Abazajian09} up to $z = 0.15$. They found systematic lower values of the specific star formation rate (sSFR, defined as the ratio between SFR and stellar mass) in filaments than in the isotropic infall region, thus providing evidence that galaxies infalling alongside filaments have experienced stronger environmental effects than galaxies infalling from other directions. Similar analysis have been done using the VIMOS  Public Extragalactic Redshift Survey \citep[VIPERS, ][]{Guzzo14,vipers} for the early Universe in \citet[][hereafter Paper 1]{Salerno2019}, and using OmegaWINGS spectroscopic survey \citep{Gullieuszik15,Moretti17} in the redshift range $0.04 \leq z \leq 0.08$ in \citet[][hereafter Paper 2]{Salerno2020}, confirming the important role played by filaments in quenching galaxies up to $z \sim 0.9$.

Post-starburst (PS) galaxies are a peculiar class of objects that are experiencing a rapid transition from being star forming to quiescent. Approximately  $\sim 1 {\rm Gyr}$ before the observing time, 
they are supposed to have had a large burst of SF but the observed spectra show very little or no ongoing SF. The evidence of a recent burst of SF comes from the presence of strong Balmer absorption lines typical of young stars \citep{Couch1987}. \cite{Paccagnella19} found that the frequency of PS galaxies is strongly dependent on the environment; they observe an increasing fraction from individual objects, binary systems, groups and clusters, suggesting a progressive increase with the halo mass. Similar results are found in Paper 2; in particular, they found that the filamentary region shows a fraction of PS galaxies greater than the isotropic infall region.

In this paper, we study the properties of a population of simulated galaxies that are falling to clusters/groups with the aim of characterising the connection between the quenching of SF in these galaxies and the different infalling regions in which they are found. Thus, we distinguish between galaxies that are infalling into clusters/groups along filaments and those that are infalling isotropically. To identify filaments, we use a sample of massive dark matter (DM) haloes (filament nodes) from the MultiDark Planck 2 cosmological simulation \citep[\mdpl, ][]{Klypin2016}. Galaxy properties are derived from these haloes by populating them with galaxies generated from the semi-analytic model of galaxy formation \sag~\citep{Cora18}, giving place to a sample of galaxy clusters. In order to evaluate the evolution of the properties of infalling galaxies, we analyse the galaxy population at three different redshifts: $z=0 \, , z=0.65 \, , z=0.85$. We focus the analysis on the comparison between the model predictions and the observational results found in M16, Paper 1 and Paper 2. We complement the analysis by comparing the evolutionary history of galaxies residing in the different environments, i.e. filaments and isotropic infall regions, that end up with the same stellar mass at $z=0$.
 
This article is organised as follows: In Section 2, we present our data sets and our method to detect filaments; Section \ref{sect:sample} deals with our galaxy samples in different environments. In Section \ref{sect:results}, we compare the fraction of red galaxies in clusters, field, filaments and the isotropic infalling region, and its redshift evolution. We discuss the implications of our results in Section \ref{sect:disct}. Finally, we summarise our main results in this section as well. 

\begin{figure*}
\includegraphics[width=15 cm]{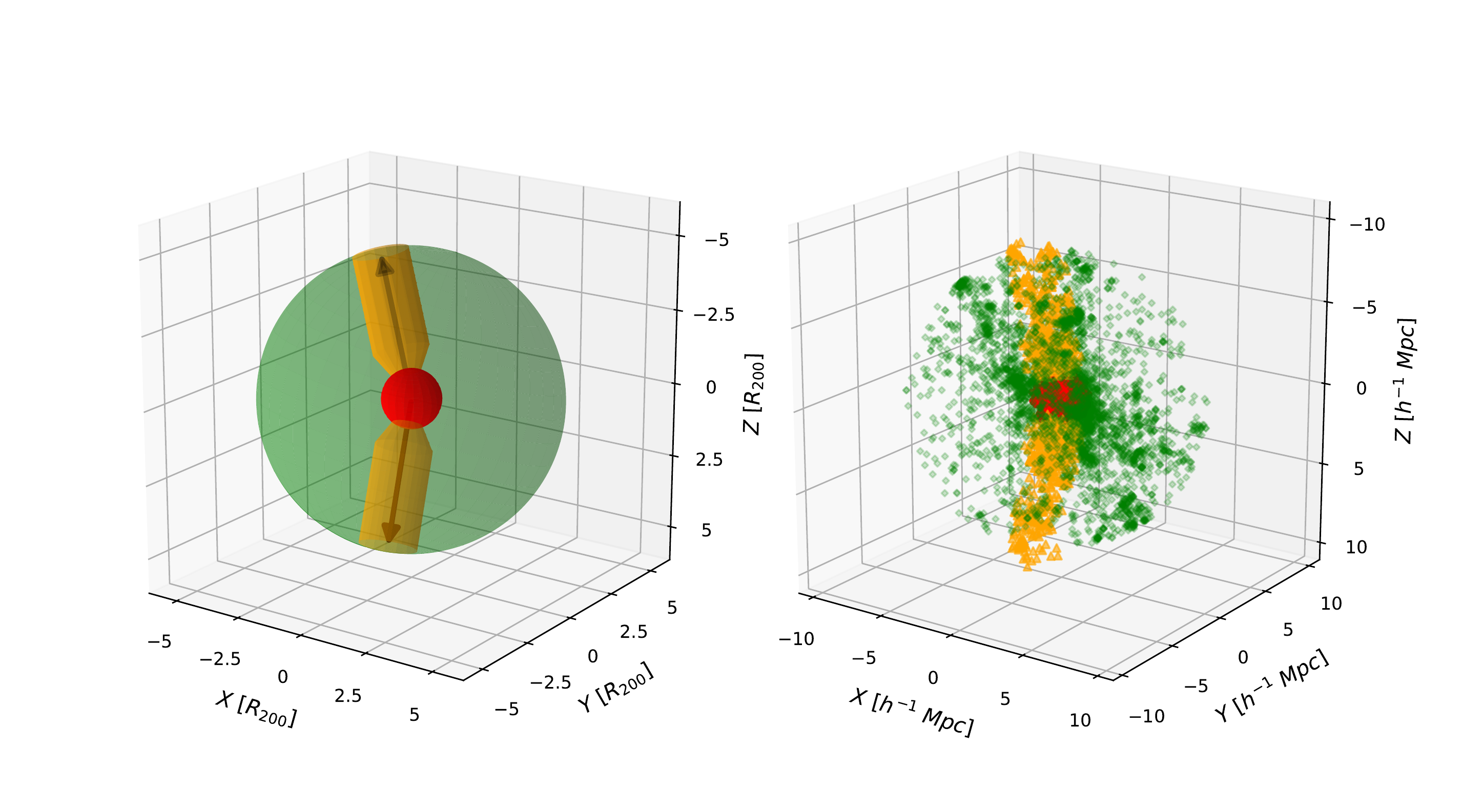}
\caption{
Left panel: scheme of the different regions found in the outskirts of a cluster (represented by a red sphere) in units of $R_{200}$: filamentary infall regions
(orange structures; arrows indicate the direction towards the corresponding filaments) and isotropic infall regions (remaining green regions of the larger sphere). Right panel: 
an example extracted from simulated results, where cluster members are shown as filled red circles, isotropically infalling galaxies are shown as green diamonds and orange triangles represent galaxies in the filamentary infall region.
}
\label{fig:regions}
\end{figure*}

%-------------------------------------------------------------------------

%%%%%%%%%%%%%%%%%%%%%%%%%%%%%%%%%%%%%%%%%%%%%%%%%%%%%%%%%%%%%%%%%%%%%%
\section{Simulated galaxy clusters and their outskirts}
\label{sect:sag}

We simulate the formation and evolution of galaxies within clusters and in their surroundings by applying the semi-analytic model of galaxy formation \sag~to the haloes and subhaloes (and their corresponding merger-trees) extracted from the \mdpl~cosmological simulation. In the following, we briefly describe the main aspects of both the \mdpl~simulation and the \sag~model. Then, we present the sample of DM
haloes selected to construct the set of galaxy clusters and explain the procedure followed to identify filaments, which allow us to define two infall regions for further analysis.

\subsection{The \mdpl~simulation}

The \mdpl~is part of the suite of DM cosmological simulations with Planck cosmologies of the \textsc{MultiDark} project \citep{Riebe13, Klypin2016}, publicly available at the \textsc{CosmoSim} database\footnote{https://www.cosmosim.org/}. The simulation follows the evolution of $3840^3$ DM
particles in a comoving box of $1 \,h^{-1}$ Gpc on a side, from an initial redshift $z=120$ to $z=0$. 
The cosmological parameters are consistent with Planck measurements of a flat $\Lambda$CDM model \citep{Planck14}: $\Omega_{\rm m}=0.307$, $\Omega_{\Lambda} = 0.693$, $\Omega_{\rm b} = 0.048$, $\sigma_8 = 0.823$, $n = 0.964$, and $h = 0.678$. The simulation counts with several halo catalogues derived from different post-processing techniques. In particular, we use the haloes catalogue and merger-trees constructed with \textsc{Rockstar} \citep{Behroozi13} and \textsc{ConsistentTrees} \citep{Behroozi13b} codes, respectively. This catalogue can be found as \textsc{mdpl2.Rockstar}\footnote{doi:10.17876/cosmosim/mdpl2/006} in the database. The outputs of the simulation are stored in $126$ snapshots between $z = 17$ and $z=0$.

\subsection{The SAG model}

The semi-analytic model of galaxy formation and evolution \sag~has its roots in the Munich model  \citep{Springel01}, and has been highly developed and improved as described in \citet{Cora06}, \citet{Lagos08}, \citet{Tecce10}, \citet{Orsi14}, \citet{MunozArancibia15},  \citet{gargiulo15}, and \citet{Cora18}. The model includes the main physical processes relevant to galaxy formation and evolution such as radiative cooling of hot gas (both in main haloes and subhaloes), quiescent SF, starbursts generated by disc instabilities and galaxy mergers
that contribute to form both a stellar bulge and a super massive black hole,
feedback from both supernovae explosions and AGN, and a detailed treatment of chemical enrichment. 
These processes regulate the circulation of baryons between different galactic components: hot gas halo, cold gas disc, stellar disc and bulge, reservoir of ejected mass. The gas phases are identified by the way in which they are generated, i.e. no temperature threshold is assumed to separate the hot gas from the cold gas. Thus, when a DM halo is identified for the first time, we assume that baryons are shocked and heated during their cosmic accretion onto the DM halo, and the mass of hot gas is computed from the baryon fraction of its virial mass. Once a hot gas halo is formed, a pressure-supported disc of cold gas is generated as a result of gas cooling. Quiescent SF takes place in this cold gas disc. 
As the galaxy evolves, the corresponding hot gas halo is redefined at each snapshot of the DM simulation by substracting from the baryon fraction of the virial mass the mass of cold gas, stars and super massive black hole of the galaxy itself and of surrounding satellite galaxies contained in the main host DM halo;
since the hot gas of galaxies is gradually removed by environmental effects when they become satellites, the hot gas mass of satellite galaxies is also discounted from the estimation of the hot gas of the main host halo at each snapshot \citep[see eq. 2 of][]{Cora18}.
The modelling of environmental effects, namely RPS and TS, are relevant for the purpose of the current study. The implementation of RPS is based on a new analytic fitting profile that model the ram pressure exerted over satellite galaxies in different environments and epochs \citep{VegaMartinez22}.
\sag~considers orphan galaxies to track the remnant satellites of haloes that are not longer resolved by the underlying simulation, whose positions and velocities are consistently calculated using an analytical model that considers dynamical friction and mass-loss by TS
\citep{Delfino:2022}\footnote{ In the current work, we use a previous version of the orbital evolution code that assumes an isothermal sphere to describe the mass profile of both the host halo and unresolved subhaloes instead of the NFW profile adopted by \citet{Delfino:2022}.}. 
%The semi-analytic model also includes the modelling of environmental effects, namely RPS and TS, which are relevant for the purpose of the current study. The implementation of RPS is based on a new analytic fitting profile that model the ram pressure exerted over satellite galaxies in different environments and epochs \citep{VegaMartinez22}.

The work of \citet{Cora18} presents the results of the dependence of the fraction of quenched MultiDark-\sag~galaxies as a function of stellar mass, halo mass and halo-centric distance for two versions of the model. They only differ 
in the dependence of the efficiency of supernova feedback on redshift. The amount of reheated and ejected mass produced by supernova explosions is proportional to the number of supernovae arising in each star-forming event and the energy released by a single supernova, and inversely proportional to the square of the virial velocity of the host (sub)halo of the corresponding galaxy, which is a measure
of its potential well. Both the estimations of the reheated and ejected mass (which are respectively transferred from the cold gas reservoir to the hot gas halo and to a reservoir of ejected mass for a later gradual reincorporation) involve efficiencies that are free parameters of the model. They also include a dependence on redshift through the factor $(1+z)^{\beta}$, being $\beta$ another free parameter of the model (see their Eqs. (10) and (12)).
In this work, we apply the version of \sag~characterised by $\beta=1.3$, which allows to recover the observed fraction of $z=0$ quenched galaxies (see their fig. 11).

\subsection{Sample of cluster-size haloes}

The set of galaxy clusters is constructed by selecting cluster-size DM haloes from the \mdpl~simulated volume. We select all the haloes at $z=0$ with mass $M_{200} \geq 10^{15} \Msol$, computed within the spherical region that encloses $200$ times the critical density, characterised by a radius $R_{200}$. We exclude from our analysis those haloes undergoing a major merger. This is done by restricting the analysis to objects that have no companion haloes more massive than $0.1\times M_{200}$, within $5 \times R_{200}$. Our selection results in a set of $34$ of the $85$ haloes more massive than $10^{15} \Msol$ in the \mdpl~volume.

Selecting the most massive haloes at $ z = 0 $ guarantees that their progenitors at $ z \sim 0.9 $ have already been formed. This criterion allows us to work with the same structures through the cosmic time within the redshift range of interest.

\subsection{Filament identification}

To identify the filamentary structures, we use the code developed by \cite{Pereyra20}. They assume that filaments are bridges of matter connecting high-density peaks.  The identification is based on the standard tools Minimal Spanning Tree (MST) and the Friends-of-Friends algorithm (FoF). The code is applied to the outputs of the \mdpl~simulation. All the DM haloes with mass greater than $5 \times 10^{11} \Msol$ are considered as nodes in the identification of the filaments.  Nevertheless, in the following analysis, we only consider the filaments that connect with any of the $34$ more massive haloes identified as described previously (first node). Therefore, the secondary node has at least a mass of $M_{200} > 5 \times 10^{11} \Msol$. For more details about the filaments identification, see \citet{Pereyra20} and \cite{Rost20}.

As a result of applying this code to \mdpl~at $z = 0$, we obtain $49$ filaments that connect with any of the $34$ DM haloes of our sample. If the algorithm is applied to the main progenitor of the $34$ DM haloes at $z = 0.64$ and $z = 0.85$, we identify  a total $89$ and $99$ filaments, respectively. It is clear that clusters are typically associated with more than one filament, being the number of filaments per node  between two and five.

\subsection{Infall regions }\label{subsec:regions}

We are  mainly interested in the properties of galaxies in the outskirts of clusters. There, we distinguish two different kinds of infall regions that are defined according to the properties of the identified filaments. We consider the direction that connects each filament, determined by the code of \citet{Pereyra20}, with their corresponding node to define a cone of revolution with an angle of $45^{\circ}$. This allows us to define two infall regions that extend beyond the limit of the cluster-size haloes, i.e. $1< R/R_{200} \le 5$, where $R$ is the distance to the cluster centre at the corresponding redshift:
\begin{enumerate}
\item The filamentary infall regions, which are volumes of space detected inside the cones defined by the filaments.
\item The isotropic infall regions, which are the remaining volumes located outside the cones.
\end{enumerate}

It is important to highlight that the general results  presented in the following sections are robust to changes in the aperture angle of the cones. It is also important to note that, as we move away from a node, the cone becomes increasingly wide. Consequently, the filamentary region can reach an excessively large size. Thereby, in order to avoid this problem, we impose a limit on the width of the defined filamentary infall regions. According to the typical sizes considered in previous works, we choose $3.4 \, h^{-1} \, \Mpc$  as the maximum width \citep{Martinez16, Salerno2019, Salerno2020}. When the base of the cone reaches this diameter, the filamentary infall region is restricted by a cylinder with the same axis as the cone. 
In the left panel of Fig. \ref{fig:regions}, we show a scheme of the different regions described above in units of $R_{200}$. The red sphere represents the cluster region. The orange areas depict the cones of revolution that continue as cylinders and represent the filamentary infall regions. Finally, the remaining parts of the sphere, painted in green, correspond to the isotropic infall region.

\begin{figure}
\begin{center}
\includegraphics[scale=0.5]{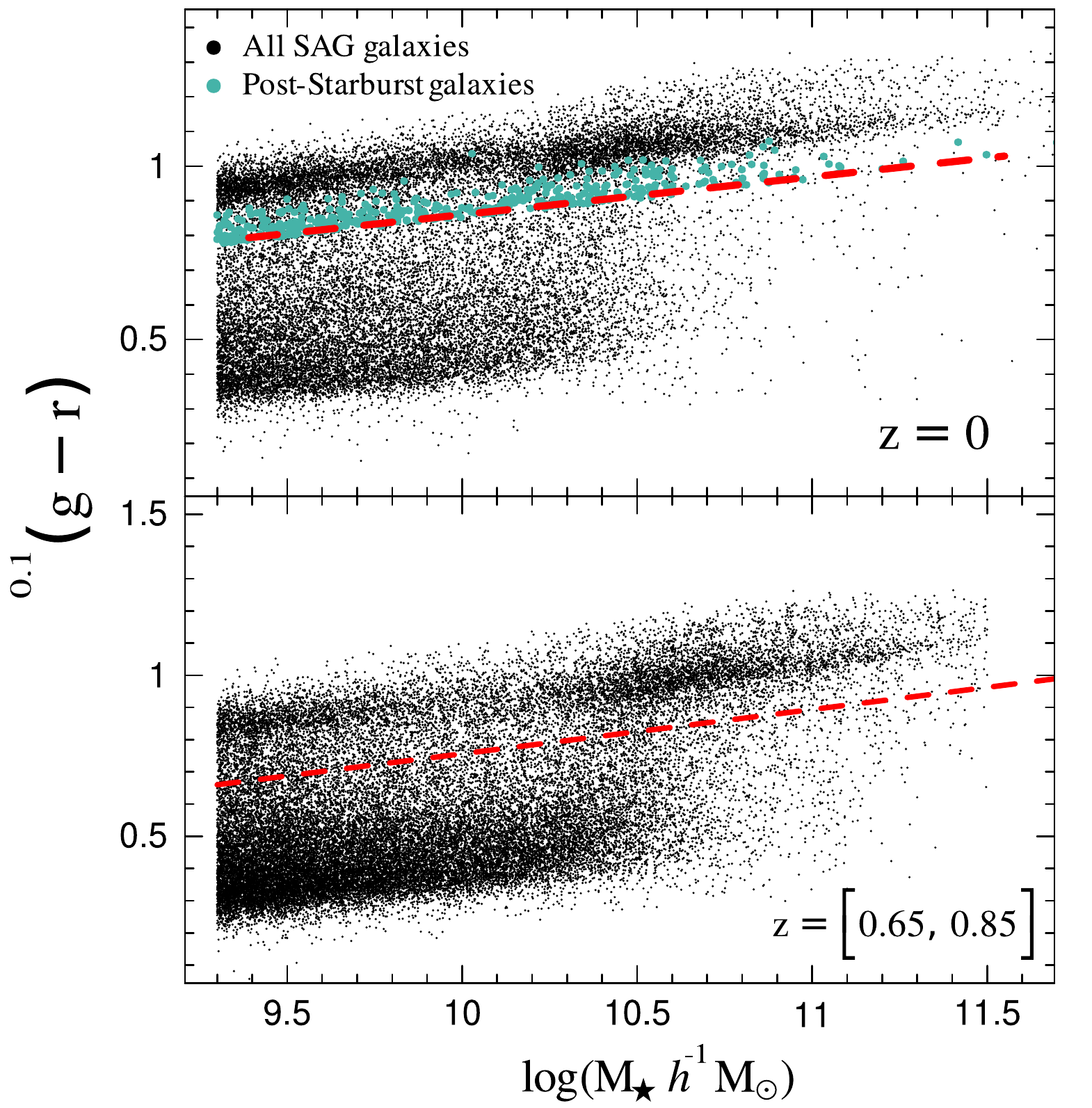}
\caption{The $^{0.1}(g-r)$ colour distributions for the 
whole sample of \sag~galaxies at $z=0$ (top panel) and at intermediate-high redshifts (bottom panel). Dashed lines indicate the adopted separations for red and blue galaxies following eq.~(\ref{ec1}). PS galaxies at $z=0$ (upper panel) are shown as turquoise points.}
\label{fig:cut}
\end{center}
\end{figure}
%-------------------------------------------------------------------------

\section{Galaxy samples}
\label{sect:sample}

To compare the predictions of the \sag~model with the observational results on the quenching of galaxies in clusters and in the infall regions, as well as the evolution with 
redshift of this phenomenon, we choose the snapshots corresponding to $z=0$, $z=0.65$ and $z=0.85$, which we will refer to as low, intermediate and high redshift bin, respectively. These choices are motivated by the mean redshifts considered in the observational studies previously described. The model predictions in the low redshift Universe will be compared with the results found in M16 and Paper 2 using the SDSS DR7 and OmegaWINGS, respectively. At intermediate and high redshift bins, model predictions will be compared with the results of Paper 1 based on VIPERS. 

For each of the $34$ DM haloes, we select all the \sag~galaxies that are at a distance $R/R_{200} \leq 5$ from the cluster centre at each redshift analysed. Thus, we construct three different galaxy catalogues for further analysis: cluster galaxies (CG), galaxies in filamentary infall regions (FRG) and galaxies in isotropic infall regions (IRG).

To be consistent with Paper 2, in this work we consider as CG all those objects that lie within the $R_{200}$ of the cluster 
(or of its main progenitor at higher $z$)
, i.e. $R/R_{200} \leq 1$. On the other hand, infalling galaxies reside either in any of the two infall regions. The right panel of Fig. \ref{fig:regions} shows an example of \sag~galaxies identified in the different regions in the space of comoving coordinates $ (X, Y, Z )$ with respect to the centre of the cluster. Cluster members are shown in red, FRG in orange, and green symbols correspond to IRG. 

Besides, we define a representative population of field galaxies to be used as a control sample. This was constructed by selecting a sub-volume of the \mdpl~simulation at $z=0$, so that the main host haloes enclosed in the region satisfy that $M_{200} \leq 2\times10^{14} \,h^{-1}\, \mathrm{M}_\odot$, and the resulting halo mass function below $10^{13} \,h^{-1}\, \mathrm{M}_\odot$ is consistent with that obtained from the complete simulation. Following these criteria, our selected region consists in a cubic volume of $\sim 43\, \mpc$ of side--length, and its most massive halo has $M_{200} = 1.63\times10^{14}\, h^{-1} \,\mathrm{M}_\odot$. 
With this selection, we ensure that all the galaxies within the cube are not gravitationally associated with large potential wells.

In all cases, we use the halo merger trees to extract the complete set of progenitors of the selected haloes to define the corresponding samples at higher redshifts.

%-------------------------------------------------------------------------

\subsection{Red, blue and post-starburst galaxies}
\label{pas_PS}

To study how the environment and the stellar mass affect the SF in galaxies, we use the  $^{0.1}(g-r)$ colour to classify galaxies %according to 
which characterises their stellar population.
The use of colour allows us to make a direct comparison with observational results like those presented in Paper 1. We use a band shift to a redshift $z=0.1$ to approximate the mean redshift of the main galaxy sample of SDSS \citep{York:2000}. Dividing the sample in five bins of stellar mass, we use the $^{0.1}(g-r)$ colour distributions to separate the population of galaxies in red and blue. Red galaxies at $z=0$ are those with \begin{equation}\label{ec1} ^{0.1}(g-r) > - \, a + b \log\left(\frac{M_{\star}}{M_{\odot}}\right), \end{equation} where $a= (0.23\pm 0.03)$ and $b=(0.11 \pm 0.03)$. The adopted threshold is shown in the upper panel of Fig.~\ref{fig:cut} where we plot the $^{0.1}(g-r)$ color-mass diagram for our sample of galaxies. For the less evolved Universe ($z=\left[0.65, 0.85 \right]$), we use slightly different parameters: $a= (1.2 \pm 0.03)$ and $b=(0.28 \pm 0.03)$, as we show in the bottom panel of Fig.~\ref{fig:cut}. The parameters of the thresholds are obtained by assuming the colour bimodality and using a method of density estimation based on gaussian mixture models \citep{Taylor15, Hahn2019}. The procedure is applied to eight  bins of stellar mass. The selection by colour  does not necessarily mean that red galaxies are passive, however, there is a very good correlation between the $^{0.1}(g-r)$ colour and the sSFR:  using a threshold of  ${\rm log}($sSFR$/{\rm yr}) = -10.7$ to separate passive from star-forming galaxies \citep{Cora18, Wetzel1012}, we find that,  $\sim 99$ per cent of the galaxies classified as red, are passive, while  $\sim 13$ per cent of the blue galaxies are in fact passive. From now on, we will use the terms red, passive or quenched galaxies indistinctly.

\begin{figure}
\begin{center}
\includegraphics[width=8.5 cm]{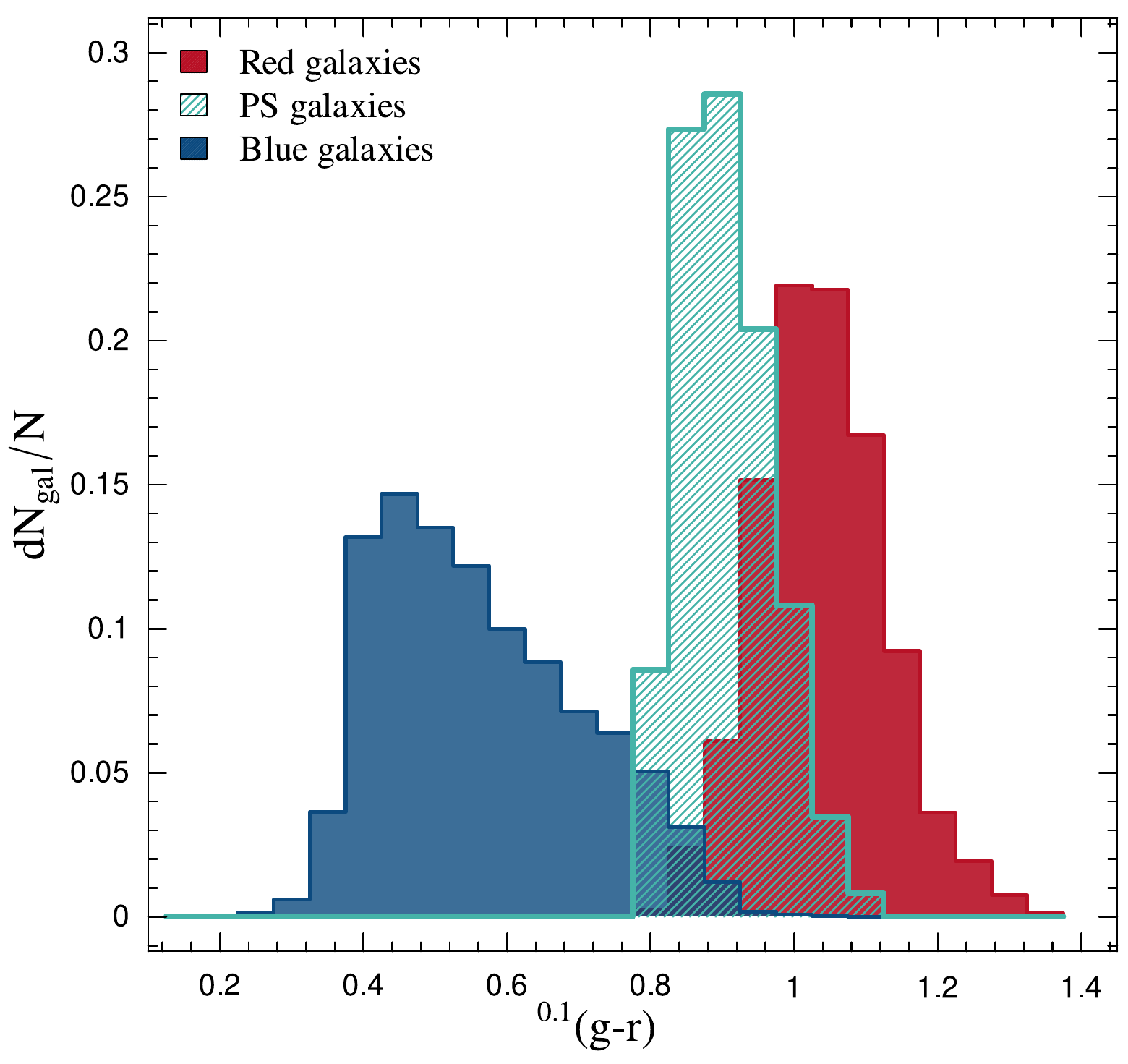}
\caption{The normalised $^{0.1}(g-r)$ colour distribution for the total sample of galaxies at $z=0$. Red and blue galaxies are selected following eq.~(\ref{ec1}) and PS galaxies are identified with the method described in Section \ref{pas_PS};  although PS galaxies are red, they have been excluded from the general sample of red galaxies.}
\label{fig:hist}
\end{center}
\end{figure}

Consistent with the observational studies previously described, PS galaxies
are only analysed in the low redshift Universe. To select PS galaxies we assume that they stopped their SF around $\sim 10^{7}$ years before \citep{Fritz2014, Paccagnella17}. Accordingly, we classify as PS
all galaxies that at $z = 0$ are classified as red, but at $z = 0.09$ (this redshift corresponds to a backward time of $\sim 10^9$ years) are blue. It is important to note that according to our selection based on colour, PS galaxies at $z=0$ are also red galaxies, however, they have been excluded from the general sample of red galaxies. In the top panel of Fig. \ref{fig:cut} we show in turquoise points the galaxies classified as PS. In Fig. \ref{fig:hist}, we present the $^{0.1}(g-r)$ colour distribution for the same galaxies as in Fig. \ref{fig:cut}, distinguishing if they are red, blue or PS. The latter ones are associated with a population in transition (e.g. \citealt{Patel09, Vulcani10, Paccagnella16}) and, therefore, it is expected that their colour distribution is between the red and the blue population. On the other hand, we see the classical colour-bimodality observed in many spectroscopic studies (e.g., \citealt{Strateva01, Hogg02, Bell04, fritz14}).

In Table \ref{tab:sample}, we show the total number of galaxies located in the different regions (CG, FRG, IRG, field galaxies), and the corresponding numbers of passive (in three redshift bins) and PS (at $z=0$) galaxies.

%-------------------------------------------------------------------------
%-------------------------------------------------------------------------

%-------------------------------------------------------------------------
\begin{table*}\centering
\begin{threeparttable}
\begin{tabular}{lccccccccccc}\toprule
           & \multicolumn{3}{c}{Total}  && \multicolumn{3}{c}{RED} && \multicolumn{2}{c}{PS}\\
            \cmidrule{2-4} \cmidrule{5-7} \cmidrule{8-10} 
            &  $z=0$ & $z=0.65$ & $z=0.85$            && $z=0$ & $z=0.65$&  $z=0.85$            &&$z=0$ &  \\ \midrule
Field Galaxies  &  1728 &  4559 & 12699 &&  392 (22\%) &  801 (17\%) & 2461 (19\%)  &&  22 (1.3\%) \\
IRG & 13469 & 8249 & 5034  && 4984 (37\%) & 1728 (20\%) & 1479 (29\%)  && 221 (1.6\%) \\
 FRG   &  2087 &  1578 & 2500  &&  801 (39\%) &  433 (27\%) &  816 (32\%) &&  42 (2.0\%) \\
CG    & 8006 &  1320 & 1823  && 4903 (61\%) & 599 (45\%) &  909 (49\%) && 202 (2.5\%)  \\
\bottomrule
\end{tabular}
\end{threeparttable}
\caption{Number of total, red and PS galaxies as a function of the environment and redshift (PS are only analysed at $z=0$). The corresponding percentages of 
red galaxies and PS galaxies with respect to the total number of galaxies in each case are shown between brackets. 
}
\label{tab:sample}
\end{table*}
%-------------------------------------------------------------------------

%-------------------------------------------------------------------------
\begin{figure*}
\begin{center}
\includegraphics[width=14 cm]{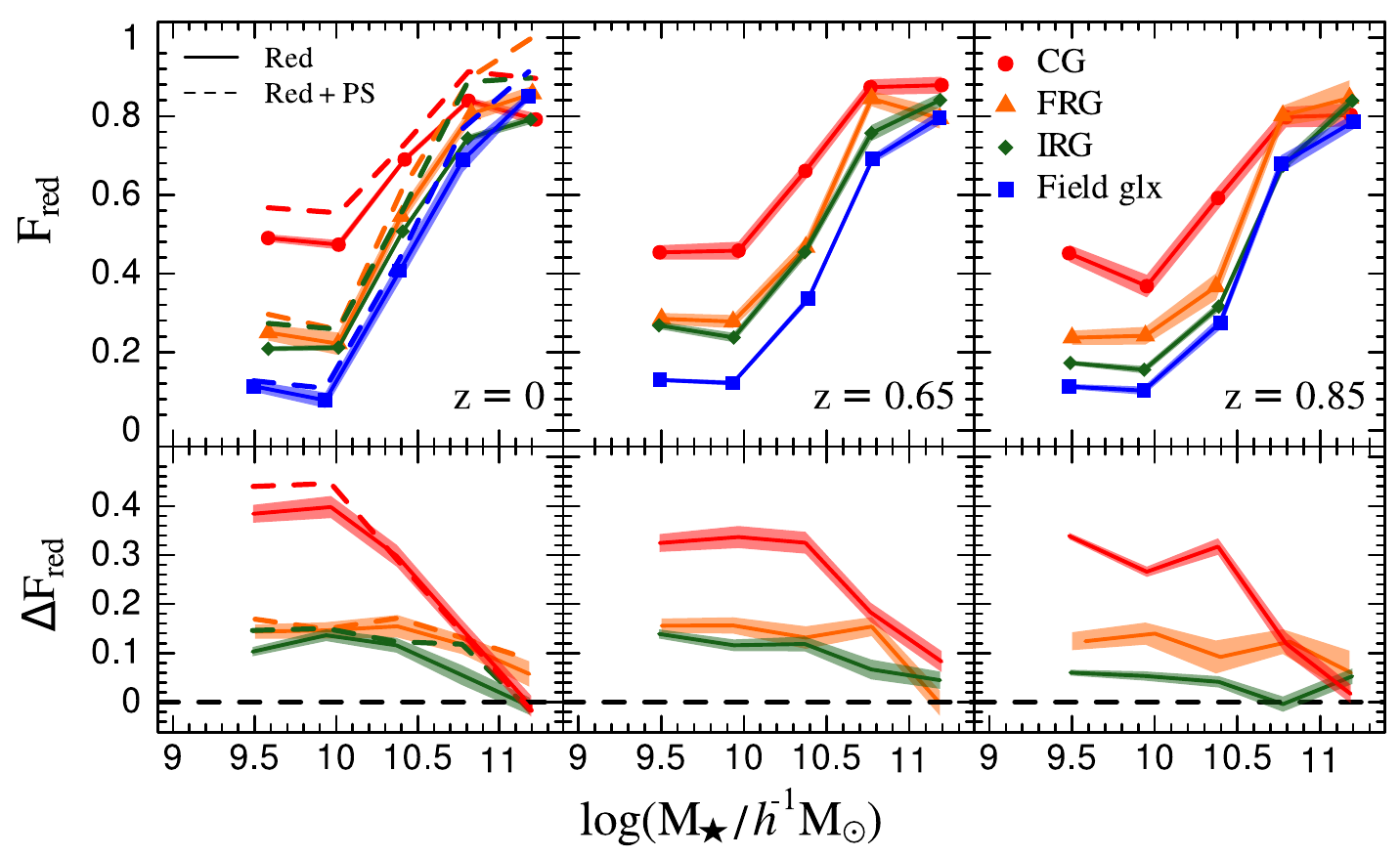}
\caption{Top panels show in solid lines the fraction of red galaxies, excluding PS galaxies, as a function of stellar mass and environment. The fractions of red
galaxies in clusters are shown as red solid circles, 
while the red fraction of FRG, IFG and field galaxies are represented by orange triangles, green diamonds, and  
blue squares, respectively. Left panels correspond to $z=0$, central panels to $z=0.5$, and right panels to $z=0.85$. At $z=0$, 
the fraction of red galaxies, including PS galaxies, are shown in dashed lines. Shaded areas represent the errors computed by using the bootstrap re-sampling technique. Bottom panels show the differences between the fractions of red galaxies relative to the fractions of field galaxies.
}
\label{color1}
\end{center}
\end{figure*}
%-------------------------------------------------------------------------

%-------------------------------------------------------------------------
\begin{figure}
    \includegraphics[scale=1.05]{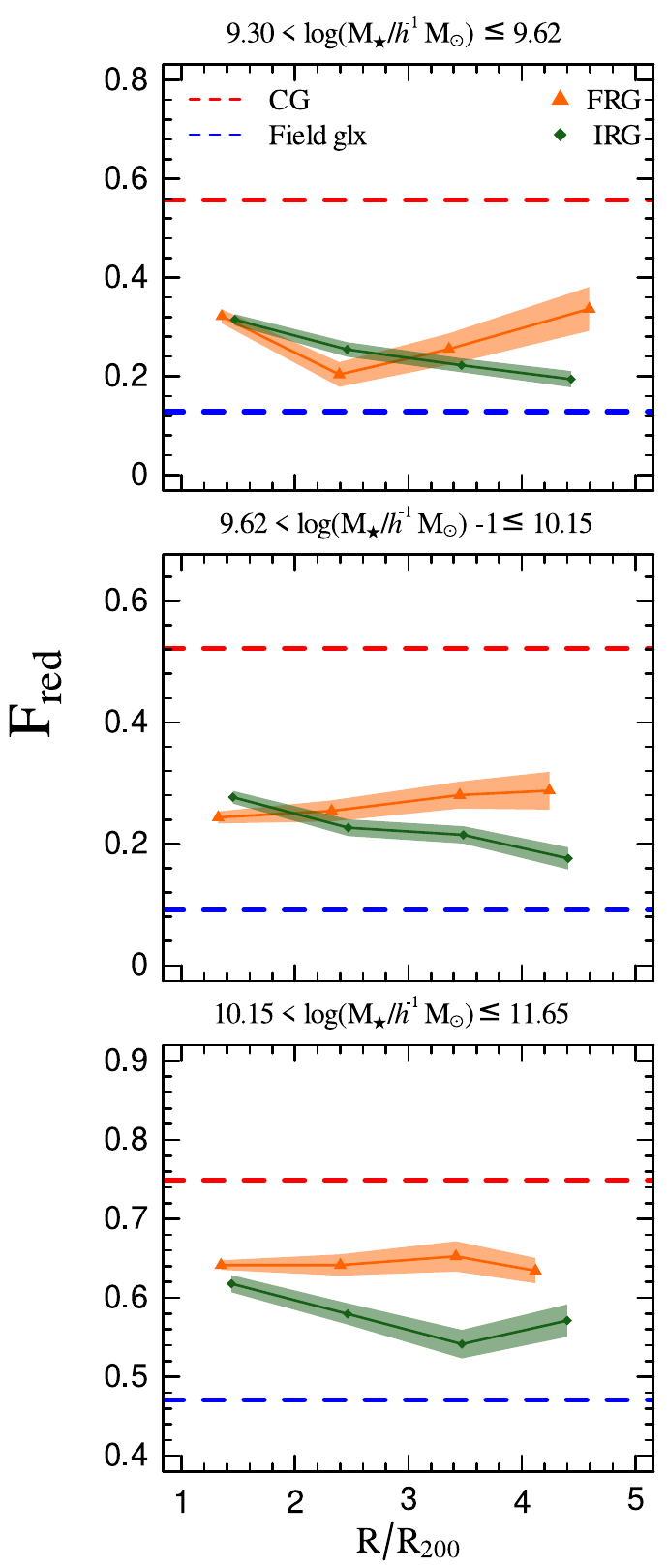}
\caption{Fraction of red FRG and IRG as a function of the normalised cluster-centric distance. Each panel corresponds to a different stellar mass bin. Error bars are calculated using the bootstrap resampling technique. Colours and symbols are as in Fig. \ref{color1}. Red and blue dashed lines indicate the mean fractions of red galaxies in clusters and in the field, respectively.}
\label{r_200}
\end{figure}
%-------------------------------------------------------------------------
%%%%%%%%%%%%%%%%%%%%%%%%%%%%%%%%%%%%%%%%%%%%%%%%%%%%%%%%%%%%%%%%%%%

\section{Results}
\label{sect:results}
\subsection{Fractions of red galaxies at different redshifts}

In the top panels of Fig. \ref{color1}, we show the fractions of red galaxies ($F_{\rm red}$) as a function of the stellar mass in the four environments considered. The left panels correspond to the local Universe ($z=0$), the central panels show the fractions at redshift $z=0.65$, and the right panels correspond to redshift $z=0.85$. Errors are computed by the bootstrap re-sampling technique. In the bottom panels, we show the residuals between each distribution relative to the field. The solid lines at $z=0$ correspond to the general sample of the red galaxies selected in Section \ref{pas_PS}, while the dashed lines show the fractions when PS galaxies are included.

The fraction of red galaxies grows with the stellar mass, and the same behaviour is observed in all the environments considered, and for all redshift bins. These results are in agreement with previous works (M16, Paper 1, \citealt{Sarron:2019}).  Moreover, the 
fraction of red galaxies in each environment differs from each other, with the exception of the one corresponding to the highest stellar mass bin ($\log{(M_{\star}/\Msol)}\sim 11.5$), for which all the environments show similar  fractions of red galaxies, independently of the redshift bin. As an overall trend, at fixed stellar mass, galaxies in clusters are more quenched than those in filamentary regions. In addition, the fractions of red galaxies in filamentary regions are larger than the fractions of red galaxies located
in the isotropic infall regions. This difference between red galaxies in filaments and in the isotropic infall region is better appreciated in the bottom panels of  Fig. \ref{color1}, being the largest difference at $z=0.85$. Finally, galaxies in the field have the lowest fraction of red galaxies, in all redshift ranges considered. For the lowest stellar mass bin, we observe that the difference between galaxies in clusters and in the field is greater at $z=0$, being the largest difference about $\sim 0.4$, as we can see in the bottom panels of Fig.~\ref{color1}. Our results suggest that the isotropic infall region is able to affect the SF in galaxies even at $z=0.85$. More interesting, galaxies that are falling into clusters along filaments undergo an extra SF quenching in comparison to those infalling isotropically.
%------------------------------------------------------------------------------------------------------------
%------------------------------------------------------------------------------------------------------------
\begin{figure*}
\begin{center}
\includegraphics[width=10cm]{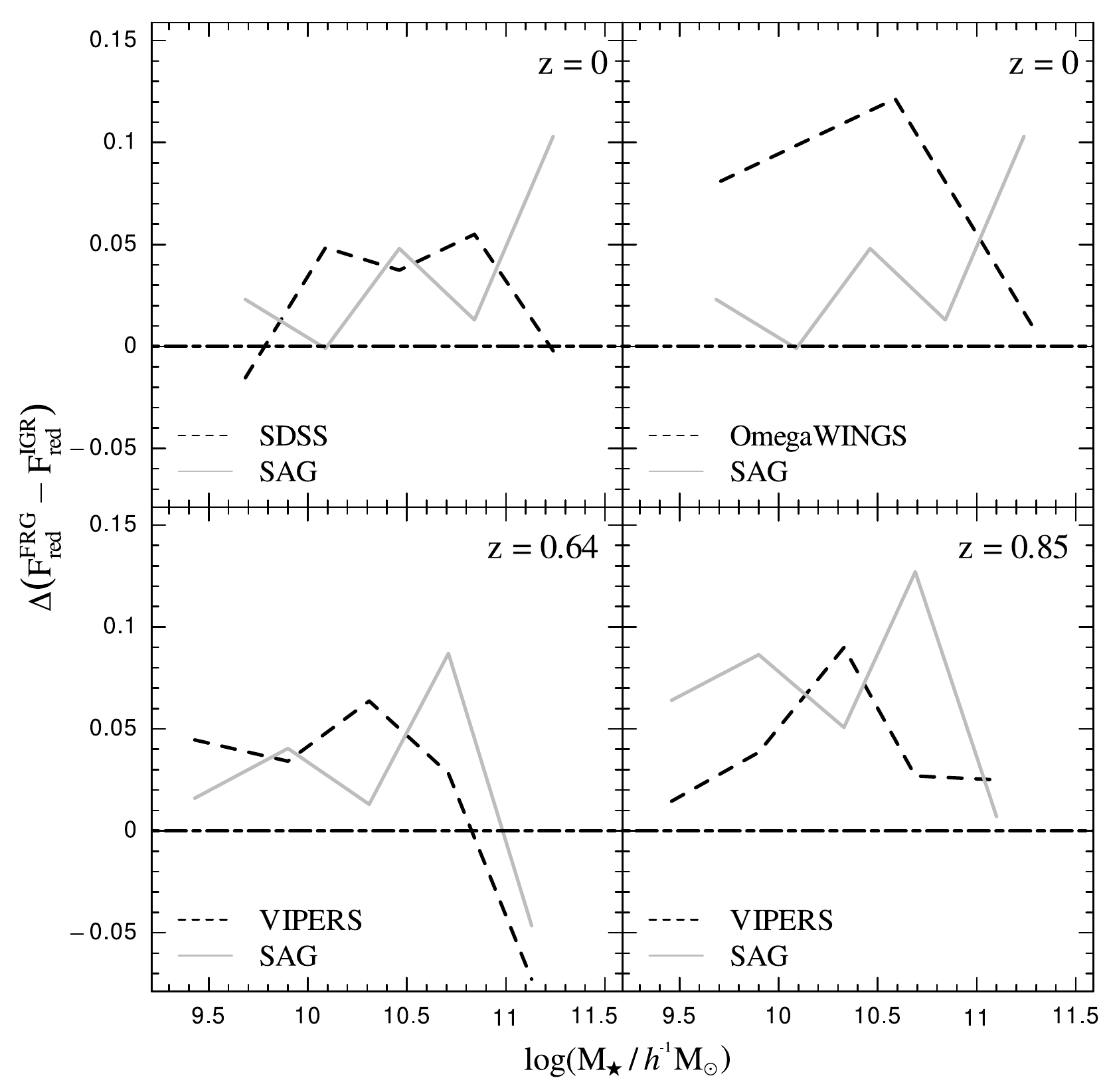}
\caption{The differences between the fractions of red FRG and red IRG predicted by \sag~(solid lines) and derived from observational data (dashed lines). The upper left and upper right panels show the comparison with SDSS (fig. 7 of M16) and with OmegaWINGS (fig. 3 of Paper 2), respectively. The bottom panels present the comparison with VIPERS at two different redshifts
(fig. 2 of Paper 1).} 
\label{dif_sag_spec}
\end{center}
\end{figure*}

%------------------------------------------------------------------------------------------------------------
%------------------------------------------------------------------------------------------------------------

\begin{figure*}
\begin{center}
\includegraphics[width=15cm]{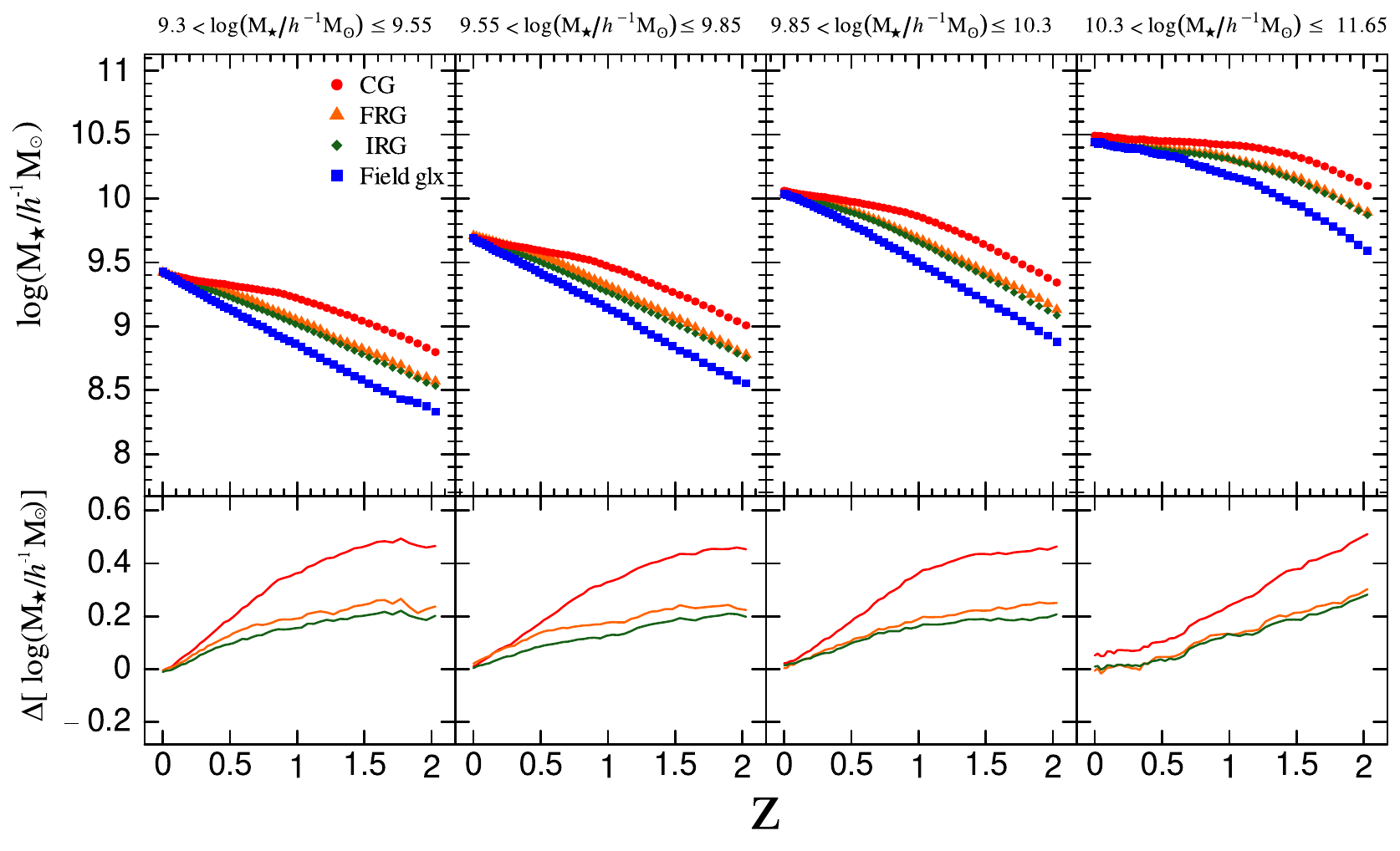}
\caption{
Meadian values 
of the stellar mass as a function of redshift for galaxies in the four selected environments. Each panel corresponds to different mass bins at $z=0$. Each point represents the median of the stellar mass for the corresponding output of the \sag~model. Bottom panels show the differences between the stellar masses of CG, FRG and IRG relative to that of galaxies in the field. Colours and symbols are as in Fig. \ref{color1}.} 
\label{mass_z} 
\end{center}
\end{figure*}

%------------------------------------------------------------------------------------------------------------
%------------------------------------------------------------------------------------------------------------

Fig. \ref{r_200} shows the fractions of red galaxies as a function of the cluster-centric distance normalised to $R_{200}$ at $z=0$. We split our sample in three bins of stellar mass: $\log(M_{\star}/\Msol)= [9.30,9.65]$, $[9.65,10.30]$ and $[10.30,11.65]$. The size of bins were chosen to have equal number of objects. The red and blue dashed lines correspond to the mean fractions of red galaxies in clusters and in the field, respectively. For the low and intermediate stellar mass bins, we observe that red galaxies in filaments and in the isotropic infall regions have fractions between the mean values of the red galaxies in clusters and in the field. For the same bins of stellar mass, the fractions of red FRG and IRG are similar for $R/R_{\textup{200}} \lesssim 3 $, while for $R/R_{\textup{200}} \gtrsim 3$ the fraction of red FRG increases considerably. In the high stellar mass bin, we observe that, for $R/ R_{\textup{200}} \gtrsim 2 $, FRG are systematically more quenched than their counterparts in the infall regions, giving place to a higher fraction of red FRG, closer to the value achieved within clusters.

\subsection{Post-starburst galaxies}

PS galaxies can be a valuable tool to analyse the impact of the infall region. Studying a sample of galaxies in the OmegaWINGS\footnote{A wide area spectroscopic survey of more than $40$ clusters of galaxies in the nearby Universe ($0.04 <z< 0.07$) that provides estimates of star formation rates, stellar masses, and measures of observed equivalent widths for the most prominent spectral lines, both in absorption and in emission \citep{Gullieuszik15}.}, \citet{Paccagnella17} find that it is composed by galaxies with different infall times, including both backsplash and virialized galaxies. As galaxies approach  $R_{200}$, processes like ram pressure could induce a burst of SF and a subsequent fast quenching, producing the characteristic PS spectrum of these galaxies. However, these mechanisms could begin at several virial radii \citep{Paccagnella17}. In Paper 2, the authors analyse the abundance of PS galaxies in clusters and in their outskirts. These authors find that the lower fraction of PS galaxies corresponds to the field and the higher one to clusters, while infall regions show intermediate values, being in the filamentary region slightly higher than in the isotropic infall region.

Here, we analyse the behaviour of PS galaxies in the \sag~model. We observe in Fig. \ref{fig:cut} and Fig. \ref{fig:hist} that the  $^{0.1}(g-r)$ colours of PS galaxies are
located between the distributions of the red and blue populations. PS galaxies are, by definition, red at $z=0$, however, the fact that they are not as red as the general population of red galaxies is a consequence of the recent SF quenching that these galaxies have experienced. As can be seen in Table~\ref{tab:sample}, the fraction of PS galaxies also depends on the environment, where the filamentary regions tend to have a larger fraction of PS galaxies than the isotropic infall region. From Table \ref{tab:sample}, we can extract the PS fractions at $z=0$ by environment: 2.5 per cent in clusters, 2.0 per cent in filamentary regions, 1.6 per cent in the isotropic infall regions and 1.3 per cent in the field. These trends are in agreement with the observational results reported in Paper 2, however, our percentages are systematically lower. This systematic shift in the PS fractions could be due to the different methods applied to select the star-forming galaxies (see the discussion in the next section).
Again, both the observational results and the predictions of the \sag~model show that the infall regions, both filamentary and isotropic ones, play an important role in the pre-processing of galaxies.

\subsection{Comparison of the predictions of the SAG model with observations}

The observational results presented in M16, Paper 1 and Paper 2 were derived using different methods to define passive galaxies and the infall regions, making a direct comparison with model predictions particularly difficult. In relation to the definition of passive galaxies,
in Paper 2 the sample of passive galaxies was obtained through the analysis of emission lines using the method originally proposed by \cite{Dressler99} and \cite{Poggianti99},
and more recently described by \citet{fritz14}. This classification uses the equivalent width of the lines  $\rm [OII]$,  $\rm [OIII]$, $\rm H_{\delta}$ and $\rm H_{\alpha}$.
On the other hand, in M16 and Paper 1, passive galaxies are selected using the  sSFR and the diagram colour-colour, respectively. In M16 the 
SFR is obtained by combining the stellar mass and the SFR likelihood distributions as outlined in Appendix
A of \citet{Brinchmann04}; passive galaxies are selected by applying a threshold in the sSFR.
In Paper 1, VIPERS galaxies are classified as passive using the $NUV-r$ vs. $r - K$ diagram of \citet{fritz14} and the code \textsc{Hyperzmass} to estimate absolute magnitudes and stellar masses. 

The observational works also present differences in the way of defining the infall regions. M16 and Paper I select pairs of groups of galaxies with a difference in the radial velocities of their baricentres, $\Delta V_{12}$, lesser than $1000\,\kms$, and a projected distance
between the baricentres, $\Delta_{12}$, smaller than $10\, h^{-1}\,\Mpc$. They consider the filamentary region as a rectangular cuboid in redshift space connecting the baricentres with a width $\Delta_{12}/2$, a length $1.5 \,h^{-1}\,\Mpc$ and $\Delta V < 1000\,\kms$.
In Paper II, the authors select all pairs of systems that are separated by less than $14\,h^{-1}\,\Mpc$ in redshift space. They consider the outskirts
of clusters as the projected volume defined by $ 1 \le R/R_{200} \le 2$, 
and $|\Delta V|\le 4\sigma$, being $\sigma$ the radial velocity dispersion of the cluster. Galaxies in filaments are those that are located within $45^{\circ}$ from the projected direction to the other node of the filament. Finally, it should be noted that in M16, Paper I and Paper II, the environments are defined in redshifts space,
while model results are analysed in 3D.

The diverse methods to select passive galaxies and to define the infall regions could explain the differences between the fractions found  
in the observational works. For example, M16 found that, for high-mass galaxies at $z=0$, all environments show passive fractions above 0.8, 
while in Paper 2 the fraction of passive galaxies is clearly below this value in clusters and
does not exceed 0.4 in the field. Our analysis, which is based on a selection by colour, allows a more direct comparison with the results presented in Paper 1. However, in order to include all the observational results in our analysis, the comparison is focused on the relative differences between the fractions of red galaxies in different
environments rather than on absolute fractions. Furthermore, we have limited the comparison to differences between the fractions of red
FRG and red IRG.

In Fig. \ref{dif_sag_spec}, we show the differences between the fractions of red FRG and red IRG
as a function of stellar mass predicted by \sag~(solid lines) and the different observational results (dashed lines): 
SDSS (M16; left top panel), OmegaWINGS (Paper2; right top panel) and VIPERS (Paper 1; bottom panels). The solid lines show the differences between the trends shown in Fig.~\ref{color1} for FRG and IRG, depicted by orange and green colours, respectively. The dashed lines were calculated from the observational data published in the corresponding articles. At $z=0$, we observe two different behaviours. As shown in  the left upper panel of Fig. \ref{dif_sag_spec}, the differences between the results from
SDSS and \sag~are small ($ \leq 0.05 $), except for the highest stellar mass bin. On the other hand, we observe larger differences and opposite trends between OmegaWINGS and \sag~(right upper panel). As already mentioned, the fractions of passive galaxies derived from OmegaWINGS data
were obtained by using methods that differ substantially from those used by M16 and those applied in the analysis of \sag~outputs, thus providing
a possible explanation for the lack of agreement between observational and simulated results. Beyond these differences, it is important to note that, for both observations and \sag~, the samples of FRG have a higher fraction of red/passive galaxies than the samples of IRG. At redshift $0.64$ and $0.85$, we found a good agreement between \sag~and VIPERS. Interestingly, for the high-mass bin at redshift $0.64$, both the model and VIPERS show fractions of passive galaxies that are larger in the IRG than in the FRG, a result that should be investigated in more detail.

\subsection{Evolution of galaxy properties}

The properties of galaxies we observe at the present epoch are the result of a variety of internal and external physical mechanisms acting throughout the entire galaxy life. Through the study of the properties of galaxies in \sag~over an extended redshift range ($0 < z < 2$), we also have the opportunity to understand the relative influence of the environments in different epochs of the Universe. 

In this section, we study the evolution of the properties of galaxies depending on the environment in which they are identified at $z = 0$. For each subsample, we analyse 
different properties of galaxy progenitors
as a function of redshift. It is important to note that we do not consider that a galaxy inhabited the same environment throughout its history. With this methodology, we want to investigate how the properties of galaxies that at $z=0$ inhabit a given environment evolve. 
We divide the galaxy samples into four stellar mass bins: $\log({\rm M_{\star}}/\Msol)=[9.30,9.55]$, $[9.55,9.85]$, $[9.85,10.30]$ and $[10.30,11.65]$. The size of the bins are chosen to have equal number of objects at $z=0$. For each bin, we track the properties of main progenitors of the selected $z=0$ galaxies across different snapshots of the simulation. Finally, for each environment and bin of stellar mass, we calculate the median of each property in all simulation outputs and analyse them as a function of redshift. We focus our analysis on the 
median values of stellar mass, hot gas, cold gas and SFR.

\subsubsection{Evolution of the stellar mass}

Fig.~\ref{mass_z} shows the evolution of the stellar mass as a function of redshift. Each point represents the median of the stellar mass for the corresponding environment and snapshot. By construction, all curves end at the same value. That is, we analyse how galaxies in each environment assembled the same amount of stellar mass. For a better visualisation, bottom panels show the residuals between the stellar mass of galaxies in clusters, in filamentary regions and in isotropic infall regions, relative to field galaxies.

We observe a similar trend in all panels of Fig. \ref{mass_z}, where main progenitors of CG show the highest stellar mass values throughout the whole redshift range probed, while field galaxies are characterised by the fastest stellar mass growth. In other words, galaxies that at $z = 0$ are in clusters assembled most of their stellar mass much earlier than galaxies in the other environments. In particular, IRG and FRG show intermediate values between those for galaxies in the field and within clusters. For the first three stellar mass bins, the progenitors of FRG are slightly, but systematically, more massive than IRG. Only in the highest stellar mass bin, the trends remain indistinguishable throughout the entire redshift range.

\begin{figure*}
\centering
\includegraphics[width=14cm]{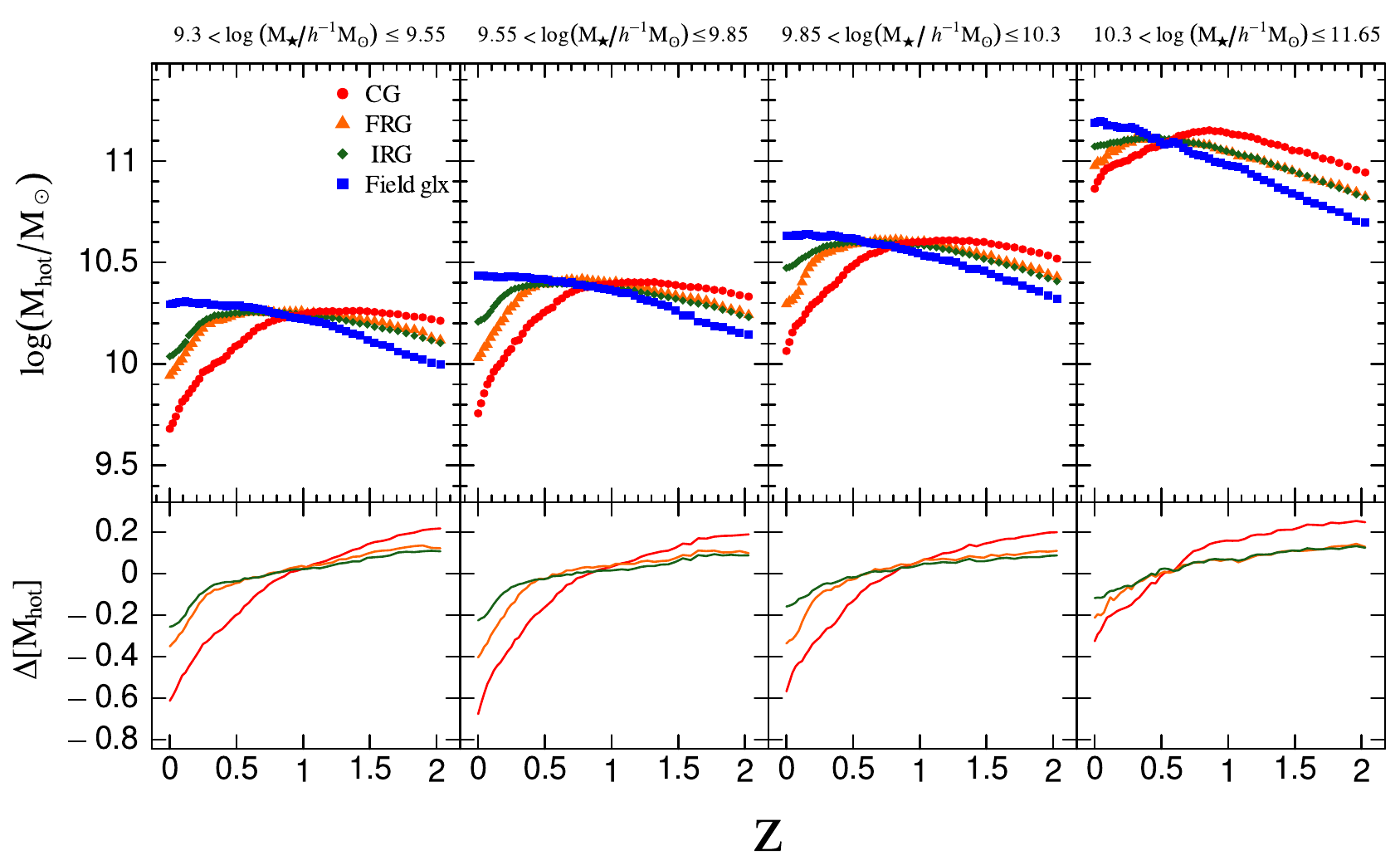}
\caption{Median values of the mass of the halo hot gas as a function of redshift for galaxies in the four selected environments in bins of stellar mass as in Fig.~\ref{mass_z}. At the bottom of each panel, the differences between the mass of hot gas of galaxies in denser environments relative to the mass of hot gas of field galaxies are shown.
}
\label{hot}
\end{figure*}

\begin{figure}
%\begin{center}
\centering
\includegraphics[width=7.5cm]{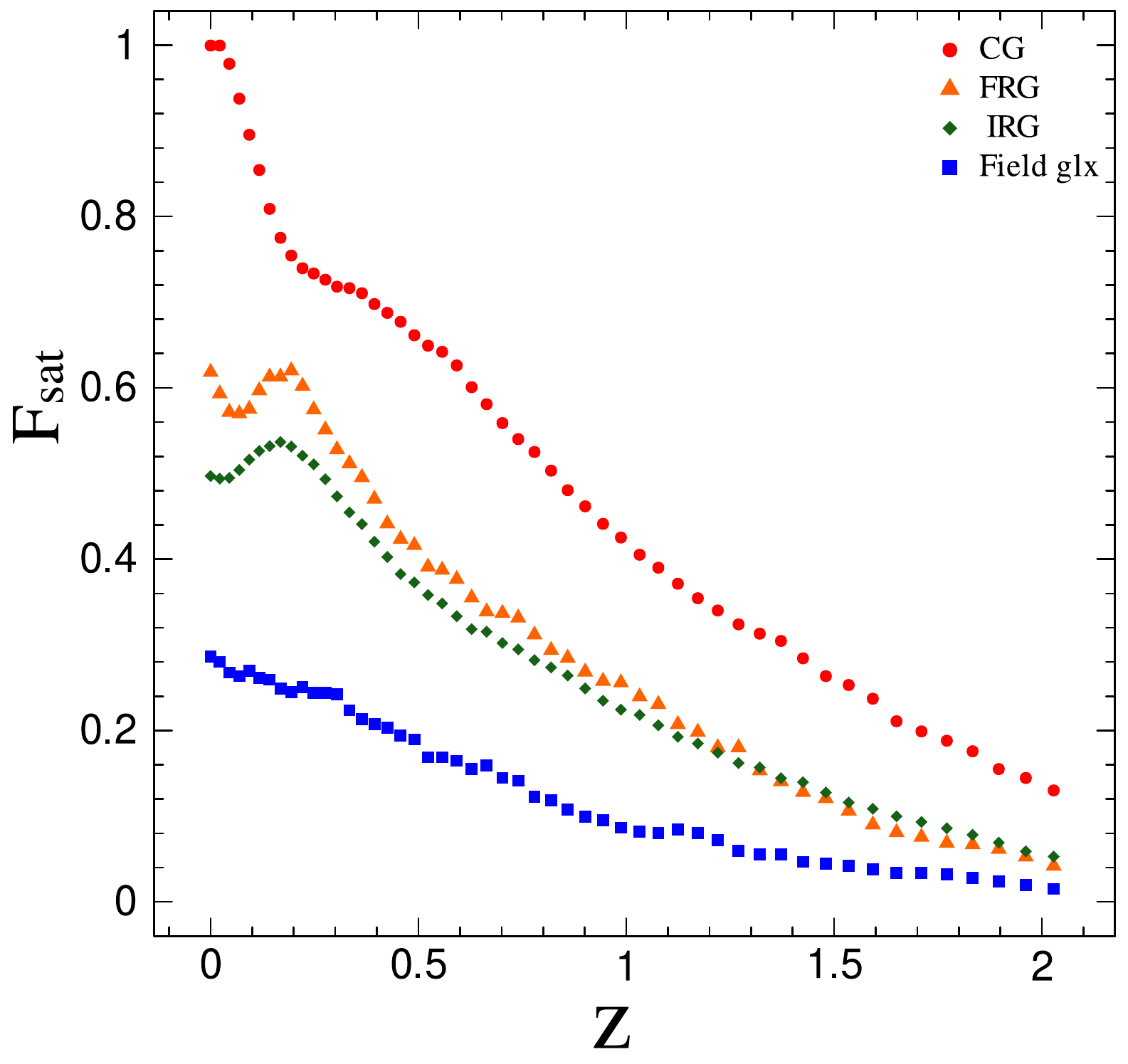}
\caption{raction of the main progenitors
of CG, FRG, IRG and field galaxies identified at $z=0$ that are classified as satellites at different epochs ($F_{\rm sat}$).
}
\label{evo:fracsat}
%\end{center}
\end{figure}

\subsubsection{Evolution of the hot and cold gas mass}

The rate of stellar mass growth is a direct consequence of the evolution of the SFR, which results from the complex circulation of mass between hot and cold gas reservoirs regulated by internal and external processes. 
Then, we focus first on the evolution of the hot gas mass.
Top panels of Figs.~\ref{hot} present the evolution with redshift of the median values of the mass of hot gas in the halo, $M_{\textup{hot}}$, of galaxies residing in different environments for the same stellar mass bins as in Fig.~\ref{mass_z};
bottom panels show the difference between the hot gas mass of galaxies in denser environments relative to the hot gas mass of field galaxies. 
In Fig.~\ref{hot} we observe that, regardless of stellar mass, the hot gas content of field galaxies grows monotonically up to the present time (with a very slight decrease near $z=0$). However, the hot gas mass of galaxies residing in clusters and in infall regions increases monotonically only up to $z\sim 1$ for the former and up to $z\sim 0.5$ for the latter, epochs in which their hot gas content begins to decrease, on average. Besides, while at earlier epochs CG have more hot gas than FRG and IRG, this trend is inverted during the decreasing stage, and the rate of decrease becomes more pronounced as we consider IRG, FRG and CG, in that order. These trends can be understood in terms of the number of central and satellite galaxies among the progenitors of $z=0$ galaxies residing in different environments, and the action of mass and environmental SF quenching on these galaxy populations.

Fig.~\ref{evo:fracsat} shows the fraction of the main progenitors of CG, FRG, IRG and field galaxies that are classified as satellites at different epochs ($F_{\rm sat}$).
Here, we do not separate the galaxies according to their $z=0$ stellar mass since general patterns are similar for galaxies in any $z=0$ stellar mass bin.
The general trends observed for any of the environments considered is that
the number of satellites increases quite monotonically as we approach to the present, simply indicating that galaxies that are satellites at a given epoch where central galaxies at some time in the past, being the number of satellites larger in denser environments at a given redshift.
This general behaviour is consistent with the structure growth in a $\Lambda$CDM cosmology. 
Many satellites found in higher density regions have been probably satellites of smaller accreted structures.
This is supported by the results obtained by \cite{Benavides20} 
from the analysis of the cosmological hydrodynamical simulation Illustris; they find that many surviving cluster galaxies at $z = 0$ are accreted as part of groups, where the percentage depends on the identification of group satellites, either given by the definition involved in the halo finder ($\sim 38$ per cent) or by considering galaxies within the virial radius of the infalling group ($\sim 20$ per cent). 
Thus, the large rate of increment of $F_{\rm sat}$ in clusters shown in Fig.~\ref{evo:fracsat} is a result of a combination of satellites accreted directly from the field (central galaxies) and both central and satellite galaxies of accreted groups.
Apart from this general increasing trend of $F_{\rm sat}$ regardless of the environment, we can see peaks and valleys at $z\lesssim 0.3$ for galaxies in the infall regions. These features could be related to the definition of a galaxy as satellite or central according to their gravitational binding energy to a given structure (cluster, infalling group) in the outskirts of clusters during the most recent stage of structure growth. We have discarded the uncertainties related to the integration of the orbits of orphan galaxies as being responsible of these variations of $F_{\rm sat}$, since the trends are similar when we exclude them from the sample. Unravelling if the origin of these variations have a physical origin or are related to definitions inherent to the halo finder deserves a deeper study that is beyond the scope of this work.

The values of $F_{\rm sat}$ at different redshifts globally characterise the change in status of the progenitors of the $z=0$ galaxies from central to satellites according to the dynamical evolution of their environments.
The hot gas of central galaxies is fed by the cosmological gas accretion during halo-mass growth, a process that does not occur, by construction, in satellite galaxies modelled by \sag~(and generally not taken into account in semi-analytic models). Hence, while $F_{\rm sat}\lesssim 0.2$, it can be considered that central galaxies dominate and, therefore, the median values of the hot gas mass increase with decreasing redshift, thus explaining the quasi-monotonic increase of the mean values of the hot gas content up to $z=0$ of field galaxies and the change of regime at $z\sim 1$ for CG and $z\sim 0.5$ for FRG and IRG, when the hot gas mass start to decrease, as shown in Fig.~\ref{hot}, according to the transformation of central galaxies in satellites. These trends are a result of a competition of both internal and external processes (the latter only act on satellite galaxies) and the dependence of their efficiencies on stellar mass. 
The hot gas mass is reduced by gas cooling in both central and satellite galaxies, a process regulated by the action of AGN feedback. The implementation of AGN feedback in \sag~is limited to the so-called `radio mode feedback' \citep{Croton06},
which is triggered by hot gas accretion onto super massive black holes, generating jets and bubbles that reduce or suppress gas cooling.
Since this effect is stronger for more massive galaxies, the median values of hot gas mass of galaxies in different environments progressively increase as we consider more massive galaxies; moreover, the amount of hot gas kept by the most massive galaxies is considerably larger than that of less massive galaxies, and decreases in a milder way towards $z=0$ for any environment. The relative difference of the hot gas content of galaxies in different environments during the decreasing phase can be explained by the action of RPS on satellite galaxies\footnote{The environmental effects (as RPS) acting on satellites of infalling groups are considered as pre-processing.}. At a given redshift, the fraction of satellites among the population of galaxy progenitors becomes larger as we consider IRG, FRG and CG (see Fig.~\ref{evo:fracsat}).
The fact that RPS is milder for more massive satellites explains the largest values of hot gas of the most massive galaxies, since neither gas cooling nor RPS contribute significantly to their decrement.
Although reheated mass by supernova feedback can enhance the hot gas reservoir of both central and satellite galaxies, these results show that the evolution of hot gas is mainly determined by the relative role of AGN feedback and RPS \citep[see][for a more detailed analysis and discussion]{cora19}.

\begin{figure*}
\centering
\includegraphics[width=14cm]{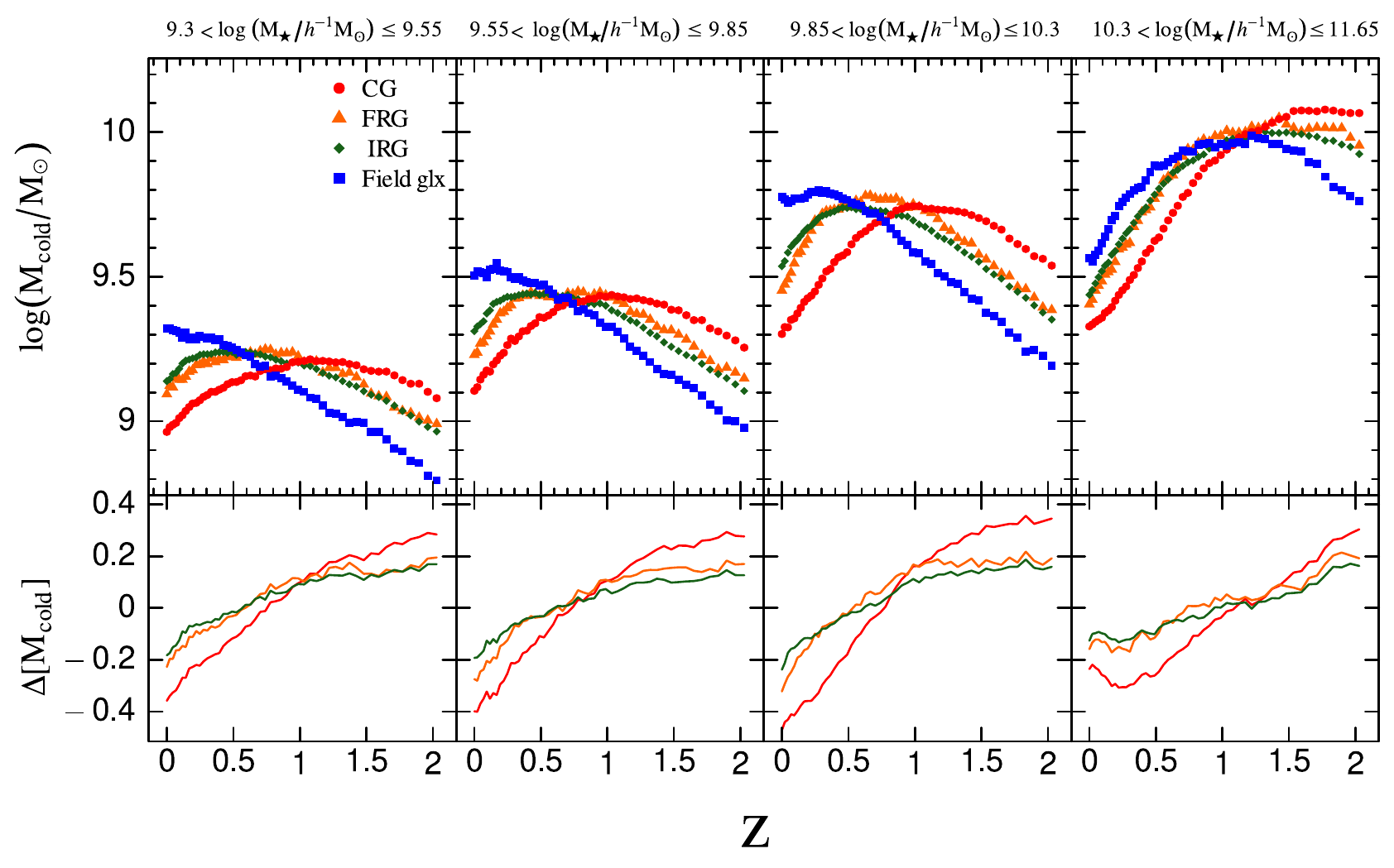}
\caption{
Median values of the mass of cold gas in the disc as a function of redshift for galaxies in the four selected environments in bins of stellar mass as in Fig. \ref{mass_z}. At the bottom of each panel, the differences between the mass of cold gas of galaxies in denser environments relative to the mass of cold gas of field galaxies are shown.
}
\label{gas}
\end{figure*}

The way in which the cold gas content in the disc evolves is directly linked to the evolution of the hot gas mass and can also be explained by the dependence on stellar mass of RPS and AGN feedback. 
Fig.~\ref{gas} is equivalent to Fig.~\ref{hot} but shows the evolution with redshift of the median values of the mass of cold gas in the disc, $M_{\rm cold}$.
The evolutionary patters of cold gas mass for different environments follow the evolution of the hot gas mass shown in Fig.~\ref{hot} for the progenitors of galaxies with $z=0$ stellar mass ${\rm log}M_{\star}/h^{-1}\,M_{\odot} \lesssim 10.3$ (first three stellar mass bins). This is consistent with the fact that the gas cooling rate is larger for higher available amounts of hot gas. However, this breaks down for the most massive galaxies because the AGN feedback reduce the gas cooling rate and prevents the cold gas reservoir from increasing despite the fact that there is a large amount of hot gas. This is particularly evident for galaxies in the field that experience a dramatic drop of their cold gas content since $z\sim 1$ which does not occur for galaxies with lower stellar mass. The significant reduction or suppression of the gas cooling rate is also evident in the most massive galaxies residing in more dense environments as the decrement of cold gas content is more abrupt than the decrement of hot gas. On top of this, for galaxies within any stellar mass range and in any environment, the cold gas reservoir becomes progressively depleted by SF and feedback from supernova.

\subsubsection{Evolution of the star formation rate}

\begin{figure*}
\begin{center}
\includegraphics[width=15cm]{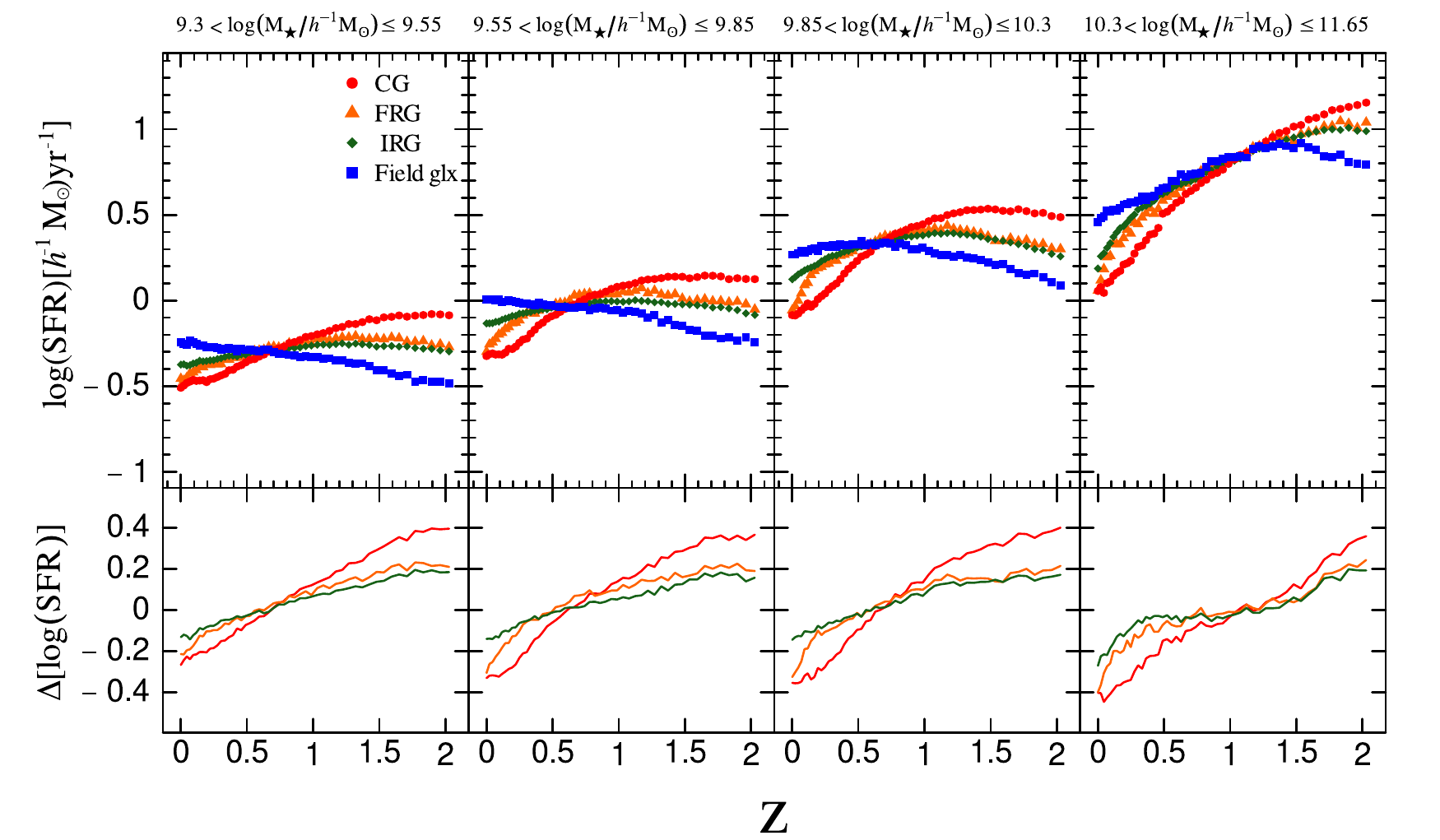}
\caption{Median values of log(SFR)
as a function of redshift for galaxies in the four selected environments in bins of stellar mass as in Fig. \ref{mass_z}. At the bottom of each panel, the differences in the SFR for galaxies in denser environments relative to field galaxies are shown. }
\label{sfr_z}
\end{center}
\end{figure*}

The evolution of the SFR predicted by \sag~is presented in Fig.~\ref{sfr_z}, where each point represents the median
of $\log({\textup{SFR}})$ for the same stellar mass bins as in Fig.~\ref{mass_z}; bottom panels show the excess of SFR of galaxies in 
denser environments relative to those in the field.
Since SF takes place from the available cold gas, the features observed in the evolution of the SFR closely follow those of the evolution of the cold gas (Fig.~\ref{gas}).
Focusing first on the general trends, we observe that, with the 
exception of field galaxies 
and progenitors of FRG and IRG at $z\gtrsim 1$, the SFR 
of the progenitors of galaxies currently residing in a given environment
increase with increasing redshift within the range of redshifts considered ($z\leq 2$),
which is consistent with observations where the highest values of SFR are at 
$z\gtrsim 1.5$ (e.g. \citealt{Lilly96, Madau14}). 

The SF in progenitors of the most massive CG
continuously falls as redshift decreases. For the rest of the mass bins, the SFR of CG progenitors has little change between redshift $\sim 2$ and $\sim 1$. After $z\sim 1$ and regardless of the stellar mass, the SFR falls more steeply and do it faster than in the infall regions.
This behaviour indicate that the majority of the main progenitors of objects classified as CG at $z = 0$ inhabited environments characterised by high levels of SF beyond $z \gtrsim 1$. The predictions of the model are in agreement with the archaeological downsizing (\citealt{Kodama04,Juneau05}), whereby more than $20-50$ per cent  of the cosmic 
SF occurred in the progenitors of CG at $z\sim 2$ (\citealt{Madau14, Chiang17}). 
The evolution of field galaxies shows a behaviour opposite to those in clusters, going from being galaxies
with the lowest SFR at $z \gtrsim 0.9$ to the highest one at $z \lesssim 0.8$. Finally, IRG and FRG show intermediate values of SFR between galaxies in the field and in clusters, in the entire redshift range. The SFR in the isotropic infall region and the filamentary region are nearly indistinguishable from each other throughout
the redshift range analyzed, with the possible exception at low redshift ($z \lesssim 0.5$).
More specifically, for $z \gtrsim 0.5$, galaxies in the filamentary region systematically form slightly more stars than galaxies in the isotropic infall region. This trend is reversed at lower redshifts, where the difference becomes progressively larger.

It is worth noting that, for all the stellar mass bins, there is a small redshift range where the SFR is similar for all the environments ($z\sim 0.7$ for the first three bins of stellar mass and $z \sim 1.1$ for the highest mass bin). These redshifts represent a change in the relation between the SFR and environment and is known as the reversal of the SFR-density relation\footnote{Effect observed in surveys at high redshifts, where the SFR-density relation at $z \sim 1$ shows an inverse behaviour compared to the well known relation at $z\sim 0$.} \citep{Elbaz07, Grutzbauch11, Popesso11}.
Besides, the way in which the values of SFR around $z\sim 0.7$ increases with increasing stellar mass is consistent with the main sequence followed by star-forming galaxies observed by the ESA {\it Herschel} space observatory \citep[][see their fig. 5]{Pearson18}; the median values of SFR predicted by the model are systematically lower because we are not restricting the analysis to star-forming galaxies.

The trends followed by progenitors of $z=0$ galaxies residing in different environments
can be understood in terms of the evolution of the cold and hot gas discussed in the previous section. These trends
are a result of two complementary aspects. On the one hand, the number of galaxies affected by RPS is lower at higher redshifts since the number of satellites among CG progenitors decrease with increasing redshifts (lower values of $F_{\rm sat}$ for higher $z$; Fig.~\ref{evo:fracsat}). On the other hand, the values of ram pressure at higher redshifts are lower because of the lower densities of the intergalactic gas that permeates the overdensities of galaxies detected at higher redshifts (like proto-clusters) \citep[][see their fig. 5]{Tecce10}, producing less stripping of hot gas (\citealt{Dannerbauer17}, \citealt{Cora18}).
Thus, at higher redshifts, galaxies have more hot gas available to feed the cold gas reservoir for subsequent SF. 
The role of AGN feedback in reducing the gas cooling rate gains relevance with the stellar mass growth of galaxy progenitors.
Besides, at a given redshift, some phenomena such as interaction between galaxies can lead to enhance the SF efficiency in higher density environments compared to lower-density ones (e.g. \citealt{Moore:1996, Villalobos:2014,Smith:2015}).

The levels of SFR at different redshifts determine the evolution of galaxy colours, a quantity that can be directly compared with observational results, contributing to their interpretation. The trends analysed in this Section explain clearly the way in which $z=0$ galaxies in different environments have achieved their colours, giving place to a larger fraction of red galaxies in denser environments, as shown in Fig.~\ref{color1}.
The larger values of $F_{\rm red}$ for more massive galaxies, also appreciated from this figure, is connected with the steeper slopes in the decreasing trend of the SFR with decreasing redshift as we consider more massive galaxies, both in clusters and in infall regions. Thus, when the reduction of SFR is more pronounced, galaxies have more time to become red, giving raise to larger values of $F_{\rm sat}$ at $z=0$.

\section{Summary and Conclusions}
\label{sect:disct}

In this work, we study the effects of the infall region on the quenching of galaxies in the redshift range $z=[0,2]$.  We use a sample of clusters and galaxies from the semi-analytic model of galaxy formation and evolution \sag~that uses DM haloes and subhaloes extracted from a MultiDark simulation. The main goals are i) to quantify the model predictions of the relative differences between the fractions of red galaxies in the different environments considered and ii) to compare the trends of these fractions of red galaxies with recent observational results. 

We find that the fractions of red galaxies in the \sag~model are strongly dependent on the stellar mass (Fig. \ref{color1}). In addition, both the filamentary and the isotropic infall regions show an intermediate behaviour between galaxies in clusters and in the field. These main characteristics were reported in observational works in the late Universe in M16 (left panel of their Fig.~7) and in Paper 2 (left panel of their Fig.~3) and at intermediate and high redshifts in Paper 2 (their Fig.~2). \cite{Sarron:2019} also report similar behaviours in the redshift range $0.15 < z < 0.7$. 

We show that in the whole range of redshift analysed, the quenching of the SF is stronger in the filamentary region than in the isotropic infall region. Our results suggest that, at $z=0.85$ (right panels of Fig.~\ref{color1}), both the isotropic and the filamentary regions are able to quench galaxies in nearly the whole range of stellar mass. On the contrary, in Paper 2 the authors concluded that only the filamentary region can pre-process galaxies. At intermediate redshifts, observations and \sag~show a similar behaviour, that is, the red fraction of both FRG and IRG are higher than the red fraction of field galaxies, being the fraction of red FRG higher than that of IRG. It is important to remark that in Paper 1 the authors performed an analysis on a sample of galaxies in a broad redshift range, while in the present work we analyse samples in narrower redshift ranges.

We also study the fraction of red galaxies, $F_{\textup{red}}$, as a function of the normalised distance to the cluster centre in the local Universe. For the low and intermediate stellar mass galaxies, and distances $R/R_{\textup{200}} \lesssim 3$, $F_{\textup{red}}$ of
IRG are similar to those of FRG, while for radii $R/R_{\textup{200}}> 3 $, the fractions of red galaxies
in the filamentary region increases considerably. For high-mass galaxies, we observe that from $R/R_{\textup{vir}} \sim 2 $, the FRG are systematically more quenched than IRG. It is important to emphasise that, regardless of stellar mass, galaxies at $R /R_{\textup{200}}> 3 $ that are falling towards clusters along filaments have been clearly more affected in their SF than those that are infalling isotropically. In general terms, these predictions of the model are in agreement with the observational results found by M16. 

Additionally, we select a sample of post-starburst galaxies. PS have $^{0.1}(g-r)$ colours intermediate between red and blue galaxies. Our results indicate that this type of objects are more likely to form in dense environments, especially in clusters. The infall regions show intermediate fractions between clusters and field, being the fraction of PS higher in the filaments than in the isotropic region, in agreement with the observational results of Paper 2. However, the percentages of PS galaxies in \sag~are systematically lower than in Paper 2. This  shift in the PS fractions could be due to the different methods used to quantify the star-forming
activity.

Finally, for the same set of environments, we have carried out a study on the evolution
of the properties of galaxies predicted by the semi-analytic model. The results show that the main progenitors of galaxies at $z = 0$ have different evolutionary behaviours depending on the stellar mass and environment. The \sag~model is able to reproduce the Lilly-Madau diagram \citep{Lilly96, Madau14}, the reversal of SFR-density relation (\citealt{Hwang2019, Elbaz07, Cooper07}), and the main sequence of galaxies \citep{Pearson18}. Comparing the evolution of 
the properties of the progenitors of galaxies that, at $z=0$, have the same stellar mass, our results show that the progenitors of galaxies that inhabit in clusters at  
$z=0$ are the most efficient to form stars at $z\gtrsim 1.5$. In contrast, field galaxies are the least efficient in that redshift range.
This could be attributed to the fact that in the progenitors of clusters (proto-clusters), the ram-pressure is not an efficient mechanism to remove the gas \citep{Dannerbauer17}, while the frequent galaxy-galaxy interactions in a dense environment tend to enhance the 
SF (e.g. \citealt{Moore:1996,Moore:1999, Gnedin:2003b, Smith:2015}). Moreover, at high redshifts, the progenitors of CG have the highest reservoirs of both hot and cold gas (Figs.~\ref{hot} and \ref{gas}, respectively),
since the number of satellites (which by construction of the \sag~model are affected by RPS) among CG progenitors decreases with increasing redshifts (Fig.~\ref{evo:fracsat}),
favouring the star formation in proto-clusters compared to the other environments.
Since clusters are the environments that first achieve high levels of SFR, it is also expected they
have the highest fractions of red galaxies at for any stellar mass.

Regarding the evolution of galaxies in the infall regions, our results show that the progenitors  of galaxies with mass $ \log(M_{\star}/h^{-1}\,{\rm M}_{\odot})\lesssim 10.3 $ that, at $z = 0$, inhabit in the filamentary region are slightly, but systematically, more massive than those in the isotropic infall region. Regarding the SFR, the \sag~model predicts that  FRG systematically form slightly more stars than IRG for $z \gtrsim  0.5$. This trend is reversed at lower redshifts, where the difference becomes progressively larger. 
These differences are also appreciated in the reservoirs of gas, that at low redshift decrease faster in FRG than in IRG. Even though these differences in the stellar mass growth and SF histories should have impacted in the evolution of the colour of galaxies, our results indicate that the red fraction of the progenitors of FRG and IRG are nearly indistinguishable from each other regardless  of the redshift or the stellar mass.  Our findings allow us to conclude that the precursors of FRG and IRG at $z=0$ began to differentiate from field galaxies long time before reaching the cluster infall region, and that filamentary regions are more efficient in the pre-processing of galaxies than the isotropic infall region.

In general agreement with the observational results, our analysis of  the predictions of the \sag~ model shows that the infall region of clusters play an important role in the pre-processing of galaxies along most of their evolutionary history. Moreover, both model predictions and observational results clearly show that the infall along filaments results in an extra SF quenching in comparison with isotropic infall. 

%%%%%%%%%%%%%%%%%%%%%%%%%%%%%%%%%%%%%%%%%%%%%%%%%%%%%%%%%%%%%%%%%%
\section*{Data availability}

The raw data of the semi-analytic model of galaxy 
formation \sag~will be shared on reasonable request to the corresponding author.

%%%%%%%%%%%%%%%%%%%%%%%%%%%%%%%%%%%%%%%%%%%%%%%%%%%%%%%%%%%%%%%%%%
\section*{Acknowledgements}

The authors thank the referee for useful comments and suggestions that help to greatly improve this manuscript. We would like to thank H\'ector J. Mart\'inez for helpful discussions about the filament identification.
This paper has been partially supported with grants from {\it Consejo Nacional de 
Investigaciones Cient\'ificas y T\'ecnicas} (CONICET, PIP 11220130100365CO) Argentina, 
and {\it Secretar\'ia de Ciencia y Tecnolog\'ia, Universidad Nacional de C\'ordoba}, Argentina.
This work used computational resources from CCAD\footnote{\url{https://ccad.unc.edu.ar/}}  - {\it Universidad Nacional de C\'ordoba}, which are part of SNCAD - MinCyT, Rep\'ublica Argentina.
SAC acknowledges funding from  CONICET (PIP-2876), {\it 
Agencia Nacional de Promoci\'on de la Investigaci\'on, el Desarrollo Tecnol\'ogico y la Innovaci\'on} (Agencia I+D+i, PICT-2018-3743), and {\it Universidad Nacional de La Plata} (G11-150), Argentina.
CVM acknowledges support from ANID/FONDECYT through grant 3200918, and he
also acknowledges support from the Max Planck Society through
a Partner Group grant.

%%%%%%%%%%%%%%%%%%%%%%%%%%%%%%%%%%%%%%%%%%%%%%%%%%



\bibliographystyle{mnras}
\bibliography{mnras} % if your bibtex file is called example.bib

\begin{thebibliography}{}
\makeatletter
\relax
\def\mn@urlcharsother{\let\do\@makeother \do\$\do\&\do\#\do\^\do\_\do\%\do\~}
\def\mn@doi{\begingroup\mn@urlcharsother \@ifnextchar [ {\mn@doi@}
  {\mn@doi@[]}}
\def\mn@doi@[#1]#2{\def\@tempa{#1}\ifx\@tempa\@empty \href
  {http://dx.doi.org/#2} {doi:#2}\else \href {http://dx.doi.org/#2} {#1}\fi
  \endgroup}
\def\mn@eprint#1#2{\mn@eprint@#1:#2::\@nil}
\def\mn@eprint@arXiv#1{\href {http://arxiv.org/abs/#1} {{\tt arXiv:#1}}}
\def\mn@eprint@dblp#1{\href {http://dblp.uni-trier.de/rec/bibtex/#1.xml}
  {dblp:#1}}
\def\mn@eprint@#1:#2:#3:#4\@nil{\def\@tempa {#1}\def\@tempb {#2}\def\@tempc
  {#3}\ifx \@tempc \@empty \let \@tempc \@tempb \let \@tempb \@tempa \fi \ifx
  \@tempb \@empty \def\@tempb {arXiv}\fi \@ifundefined
  {mn@eprint@\@tempb}{\@tempb:\@tempc}{\expandafter \expandafter \csname
  mn@eprint@\@tempb\endcsname \expandafter{\@tempc}}}

\bibitem[\protect\citeauthoryear{{Abadi}, {Moore}  \& {Bower}}{{Abadi}
  et~al.}{1999}]{Abadi:1999}
{Abadi} M.~G.,  {Moore} B.,   {Bower} R.~G.,  1999, \mnras, \href
  {http://adsabs.harvard.edu/abs/1999MNRAS.308..947A} {308, 947}

\bibitem[\protect\citeauthoryear{{Abazajian} et~al.,}{{Abazajian}
  et~al.}{2009}]{Abazajian09}
{Abazajian} K.~N.,  et~al., 2009, \mn@doi [\apjs]
  {10.1088/0067-0049/182/2/543}, \href
  {http://adsabs.harvard.edu/abs/2009ApJS..182..543A} {182, 543}

\bibitem[\protect\citeauthoryear{{Arag{\'o}n-Calvo}, {van de Weygaert}  \&
  {Jones}}{{Arag{\'o}n-Calvo} et~al.}{2010}]{aragon10}
{Arag{\'o}n-Calvo} M.~A.,  {van de Weygaert} R.,   {Jones} B.~J.~T.,  2010,
  \mn@doi [\mnras] {10.1111/j.1365-2966.2010.17263.x}, \href
  {http://adsabs.harvard.edu/abs/2010MNRAS.408.2163A} {408, 2163}

\bibitem[\protect\citeauthoryear{{Bah{\'e}}, {McCarthy}, {Balogh}  \&
  {Font}}{{Bah{\'e}} et~al.}{2013}]{Bahe:2013}
{Bah{\'e}} Y.~M.,  {McCarthy} I.~G.,  {Balogh} M.~L.,   {Font} A.~S.,  2013,
  \mn@doi [\mnras] {10.1093/mnras/stt109}, \href
  {http://adsabs.harvard.edu/abs/2013MNRAS.430.3017B} {430, 3017}

\bibitem[\protect\citeauthoryear{{Balogh}, {Navarro}  \& {Morris}}{{Balogh}
  et~al.}{2000}]{Balogh:2000}
{Balogh} M.~L.,  {Navarro} J.~F.,   {Morris} S.~L.,  2000, \mn@doi [\apj]
  {10.1086/309323}, \href {http://adsabs.harvard.edu/abs/2000ApJ...540..113B}
  {540, 113}

\bibitem[\protect\citeauthoryear{{Beckmann} et~al.,}{{Beckmann}
  et~al.}{2017}]{Beckmann17}
{Beckmann} R.~S.,  et~al., 2017, \mn@doi [\mnras] {10.1093/mnras/stx1831},
  \href {https://ui.adsabs.harvard.edu/abs/2017MNRAS.472..949B} {472, 949}

\bibitem[\protect\citeauthoryear{{Behroozi}, {Wechsler}  \& {Wu}}{{Behroozi}
  et~al.}{2013a}]{Behroozi13}
{Behroozi} P.~S.,  {Wechsler} R.~H.,   {Wu} H.-Y.,  2013a, \mn@doi [\apj]
  {10.1088/0004-637X/762/2/109}, \href
  {https://ui.adsabs.harvard.edu/abs/2013ApJ...762..109B} {762, 109}

\bibitem[\protect\citeauthoryear{{Behroozi}, {Wechsler}, {Wu}, {Busha},
  {Klypin}  \& {Primack}}{{Behroozi} et~al.}{2013b}]{Behroozi13b}
{Behroozi} P.~S.,  {Wechsler} R.~H.,  {Wu} H.-Y.,  {Busha} M.~T.,  {Klypin}
  A.~A.,   {Primack} J.~R.,  2013b, \mn@doi [\apj]
  {10.1088/0004-637X/763/1/18}, \href
  {https://ui.adsabs.harvard.edu/abs/2013ApJ...763...18B} {763, 18}

\bibitem[\protect\citeauthoryear{{Bekki}}{{Bekki}}{2009}]{Bekki:2009}
{Bekki} K.,  2009, \mn@doi [\mnras] {10.1111/j.1365-2966.2009.15431.x}, \href
  {http://adsabs.harvard.edu/abs/2009MNRAS.399.2221B} {399, 2221}

\bibitem[\protect\citeauthoryear{{Bell} et~al.,}{{Bell} et~al.}{2004}]{Bell04}
{Bell} E.~F.,  et~al., 2004, \mn@doi [\apj] {10.1086/420778}, \href
  {https://ui.adsabs.harvard.edu/abs/2004ApJ...608..752B} {608, 752}

\bibitem[\protect\citeauthoryear{{Benavides}, {Sales}  \& {Abadi}}{{Benavides}
  et~al.}{2020}]{Benavides20}
{Benavides} J.~A.,  {Sales} L.~V.,   {Abadi} M.~G.,  2020, \mn@doi [\mnras]
  {10.1093/mnras/staa2636}, \href
  {https://ui.adsabs.harvard.edu/abs/2020MNRAS.498.3852B} {498, 3852}

\bibitem[\protect\citeauthoryear{{Binney} \& {Tremaine}}{{Binney} \&
  {Tremaine}}{2008}]{Binney:87}
{Binney} J.,  {Tremaine} S.,  2008, {Galactic Dynamics: Second Edition}

\bibitem[\protect\citeauthoryear{{Bond}, {Kofman}  \& {Pogosyan}}{{Bond}
  et~al.}{1996}]{Bond96}
{Bond} J.~R.,  {Kofman} L.,   {Pogosyan} D.,  1996, \mn@doi [\nat]
  {10.1038/380603a0}, \href {http://adsabs.harvard.edu/abs/1996Natur.380..603B}
  {380, 603}

\bibitem[\protect\citeauthoryear{{Bower}, {Benson}  \& {Crain}}{{Bower}
  et~al.}{2012}]{Bower:2012}
{Bower} R.~G.,  {Benson} A.~J.,   {Crain} R.~A.,  2012, \mn@doi [\mnras]
  {10.1111/j.1365-2966.2012.20516.x}, \href
  {https://ui.adsabs.harvard.edu/abs/2012MNRAS.422.2816B} {422, 2816}

\bibitem[\protect\citeauthoryear{{Brinchmann}, {Charlot}, {White}, {Tremonti},
  {Kauffmann}, {Heckman}  \& {Brinkmann}}{{Brinchmann}
  et~al.}{2004}]{Brinchmann04}
{Brinchmann} J.,  {Charlot} S.,  {White} S.~D.~M.,  {Tremonti} C.,  {Kauffmann}
  G.,  {Heckman} T.,   {Brinkmann} J.,  2004, \mn@doi [\mnras]
  {10.1111/j.1365-2966.2004.07881.x}, \href
  {https://ui.adsabs.harvard.edu/abs/2004MNRAS.351.1151B} {351, 1151}

\bibitem[\protect\citeauthoryear{{Cautun}, {van de Weygaert}  \&
  {Jones}}{{Cautun} et~al.}{2013}]{Cautun13}
{Cautun} M.,  {van de Weygaert} R.,   {Jones} B.~J.~T.,  2013, \mn@doi [\mnras]
  {10.1093/mnras/sts416}, \href
  {http://adsabs.harvard.edu/abs/2013MNRAS.429.1286C} {429, 1286}

\bibitem[\protect\citeauthoryear{{Chan}, {Kere{\v{s}}}, {Wetzel}, {Hopkins},
  {Faucher-Gigu{\`e}re}, {El-Badry}, {Garrison-Kimmel}  \&
  {Boylan-Kolchin}}{{Chan} et~al.}{2018}]{Chan18}
{Chan} T.~K.,  {Kere{\v{s}}} D.,  {Wetzel} A.,  {Hopkins} P.~F.,
  {Faucher-Gigu{\`e}re} C.~A.,  {El-Badry} K.,  {Garrison-Kimmel} S.,
  {Boylan-Kolchin} M.,  2018, \mn@doi [\mnras] {10.1093/mnras/sty1153}, \href
  {https://ui.adsabs.harvard.edu/abs/2018MNRAS.478..906C} {478, 906}

\bibitem[\protect\citeauthoryear{{Chiang}, {Overzier}, {Gebhardt}  \&
  {Henriques}}{{Chiang} et~al.}{2017}]{Chiang17}
{Chiang} Y.-K.,  {Overzier} R.~A.,  {Gebhardt} K.,   {Henriques} B.,  2017,
  \mn@doi [\apjl] {10.3847/2041-8213/aa7e7b}, \href
  {https://ui.adsabs.harvard.edu/abs/2017ApJ...844L..23C} {844, L23}

\bibitem[\protect\citeauthoryear{{Colberg}, {White}, {Jenkins}  \&
  {Pearce}}{{Colberg} et~al.}{1999}]{Colberg:1999}
{Colberg} J.~M.,  {White} S.~D.~M.,  {Jenkins} A.,   {Pearce} F.~R.,  1999,
  \mn@doi [\mnras] {10.1046/j.1365-8711.1999.02400.x}, \href
  {https://ui.adsabs.harvard.edu/abs/1999MNRAS.308..593C} {308, 593}

\bibitem[\protect\citeauthoryear{{Cooper} et~al.,}{{Cooper}
  et~al.}{2007}]{Cooper07}
{Cooper} M.~C.,  et~al., 2007, \mn@doi [\mnras]
  {10.1111/j.1365-2966.2007.11534.x}, \href
  {https://ui.adsabs.harvard.edu/abs/2007MNRAS.376.1445C} {376, 1445}

\bibitem[\protect\citeauthoryear{{Cora}}{{Cora}}{2006}]{Cora06}
{Cora} S.~A.,  2006, \mn@doi [\mnras] {10.1111/j.1365-2966.2006.10271.x}, \href
  {https://ui.adsabs.harvard.edu/abs/2006MNRAS.368.1540C} {368, 1540}

\bibitem[\protect\citeauthoryear{{Cora} et~al.,}{{Cora} et~al.}{2018}]{Cora18}
{Cora} S.~A.,  et~al., 2018, \mn@doi [\mnras] {10.1093/mnras/sty1131}, \href
  {https://ui.adsabs.harvard.edu/abs/2018MNRAS.479....2C} {479, 2}

\bibitem[\protect\citeauthoryear{{Cora}, {Hough}, {Vega-Mart{\'\i}nez}  \&
  {Orsi}}{{Cora} et~al.}{2019}]{cora19}
{Cora} S.~A.,  {Hough} T.,  {Vega-Mart{\'\i}nez} C.~A.,   {Orsi} {\'A}.~A.,
  2019, \mn@doi [\mnras] {10.1093/mnras/sty3214}, \href
  {https://ui.adsabs.harvard.edu/abs/2019MNRAS.483.1686C} {483, 1686}

\bibitem[\protect\citeauthoryear{{Couch} \& {Sharples}}{{Couch} \&
  {Sharples}}{1987}]{Couch1987}
{Couch} W.~J.,  {Sharples} R.~M.,  1987, \mn@doi [\mnras]
  {10.1093/mnras/229.3.423}, \href
  {https://ui.adsabs.harvard.edu/abs/1987MNRAS.229..423C} {229, 423}

\bibitem[\protect\citeauthoryear{{Crain} et~al.,}{{Crain}
  et~al.}{2015}]{eagle_2}
{Crain} R.~A.,  et~al., 2015, \mn@doi [\mnras] {10.1093/mnras/stv725}, \href
  {https://ui.adsabs.harvard.edu/abs/2015MNRAS.450.1937C} {450, 1937}

\bibitem[\protect\citeauthoryear{{Croton} et~al.,}{{Croton}
  et~al.}{2006}]{Croton06}
{Croton} D.~J.,  et~al., 2006, \mn@doi [\mnras]
  {10.1111/j.1365-2966.2005.09675.x}, \href
  {https://ui.adsabs.harvard.edu/abs/2006MNRAS.365...11C} {365, 11}

\bibitem[\protect\citeauthoryear{{Cui} et~al.,}{{Cui}
  et~al.}{2018}]{threehundred}
{Cui} W.,  et~al., 2018, \mn@doi [\mnras] {10.1093/mnras/sty2111}, \href
  {https://ui.adsabs.harvard.edu/abs/2018MNRAS.480.2898C} {480, 2898}

\bibitem[\protect\citeauthoryear{{Dannerbauer} et~al.,}{{Dannerbauer}
  et~al.}{2017}]{Dannerbauer17}
{Dannerbauer} H.,  et~al., 2017, \mn@doi [\aap] {10.1051/0004-6361/201730449},
  \href {https://ui.adsabs.harvard.edu/abs/2017A&A...608A..48D} {608, A48}

\bibitem[\protect\citeauthoryear{{Delfino}, {Sc{\'o}ccola}, {Cora},
  {Vega-Mart{\'\i}nez}  \& {Gargiulo}}{{Delfino} et~al.}{2022}]{Delfino:2022}
{Delfino} F.~M.,  {Sc{\'o}ccola} C.~G.,  {Cora} S.~A.,  {Vega-Mart{\'\i}nez}
  C.~A.,   {Gargiulo} I.~D.,  2022, \mn@doi [\mnras] {10.1093/mnras/stab3494},
  \href {https://ui.adsabs.harvard.edu/abs/2022MNRAS.510.2900D} {510, 2900}

\bibitem[\protect\citeauthoryear{{Donnari} et~al.,}{{Donnari}
  et~al.}{2021}]{donnari_2021}
{Donnari} M.,  et~al., 2021, \mn@doi [\mnras] {10.1093/mnras/staa3006}, \href
  {https://ui.adsabs.harvard.edu/abs/2021MNRAS.500.4004D} {500, 4004}

\bibitem[\protect\citeauthoryear{{Dressler}}{{Dressler}}{1980}]{dressler80}
{Dressler} A.,  1980, \mn@doi [\apj] {10.1086/157753}, \href
  {http://adsabs.harvard.edu/abs/1980ApJ...236..351D} {236, 351}

\bibitem[\protect\citeauthoryear{{Dressler} \& {Gunn}}{{Dressler} \&
  {Gunn}}{1983}]{Dressler:83}
{Dressler} A.,  {Gunn} J.~E.,  1983, \mn@doi [\apj] {10.1086/161093}, \href
  {https://ui.adsabs.harvard.edu/abs/1983ApJ...270....7D} {270, 7}

\bibitem[\protect\citeauthoryear{{Dressler}, {Smail}, {Poggianti}, {Butcher},
  {Couch}, {Ellis}  \& {Oemler}}{{Dressler} et~al.}{1999}]{Dressler99}
{Dressler} A.,  {Smail} I.,  {Poggianti} B.~M.,  {Butcher} H.,  {Couch} W.~J.,
  {Ellis} R.~S.,   {Oemler} Augustus J.,  1999, \mn@doi [\apjs]
  {10.1086/313213}, \href
  {https://ui.adsabs.harvard.edu/abs/1999ApJS..122...51D} {122, 51}

\bibitem[\protect\citeauthoryear{{Ebeling}, {Barrett}  \& {Donovan}}{{Ebeling}
  et~al.}{2004}]{Ebeling:2004}
{Ebeling} H.,  {Barrett} E.,   {Donovan} D.,  2004, \mn@doi [\apjl]
  {10.1086/422750}, \href
  {https://ui.adsabs.harvard.edu/abs/2004ApJ...609L..49E} {609, L49}

\bibitem[\protect\citeauthoryear{{Elbaz} et~al.,}{{Elbaz}
  et~al.}{2007}]{Elbaz07}
{Elbaz} D.,  et~al., 2007, \mn@doi [\aap] {10.1051/0004-6361:20077525}, \href
  {https://ui.adsabs.harvard.edu/abs/2007A&A...468...33E} {468, 33}

\bibitem[\protect\citeauthoryear{{Fritz} et~al.,}{{Fritz}
  et~al.}{2014a}]{fritz14}
{Fritz} A.,  et~al., 2014a, \mn@doi [\aap] {10.1051/0004-6361/201322379}, \href
  {http://adsabs.harvard.edu/abs/2014A%26A...563A..92F} {563, A92}

\bibitem[\protect\citeauthoryear{{Fritz} et~al.,}{{Fritz}
  et~al.}{2014b}]{Fritz2014}
{Fritz} J.,  et~al., 2014b, \mn@doi [\aap] {10.1051/0004-6361/201323138}, \href
  {https://ui.adsabs.harvard.edu/abs/2014A&A...566A..32F} {566, A32}

\bibitem[\protect\citeauthoryear{{Gargiulo} et~al.,}{{Gargiulo}
  et~al.}{2015}]{gargiulo15}
{Gargiulo} I.~D.,  et~al., 2015, \mn@doi [\mnras] {10.1093/mnras/stu2272},
  \href {https://ui.adsabs.harvard.edu/abs/2015MNRAS.446.3820G} {446, 3820}

\bibitem[\protect\citeauthoryear{{Gnedin}}{{Gnedin}}{2003}]{Gnedin:2003b}
{Gnedin} O.~Y.,  2003, \mn@doi [\apj] {10.1086/374774}, \href
  {https://ui.adsabs.harvard.edu/abs/2003ApJ...589..752G} {589, 752}

\bibitem[\protect\citeauthoryear{{G{\'o}mez} et~al.,}{{G{\'o}mez}
  et~al.}{2003}]{gomez03}
{G{\'o}mez} P.~L.,  et~al., 2003, \mn@doi [\apj] {10.1086/345593}, \href
  {http://adsabs.harvard.edu/abs/2003ApJ...584..210G} {584, 210}

\bibitem[\protect\citeauthoryear{{Gr{\"u}tzbauch}, {Chuter}, {Conselice},
  {Bauer}, {Bluck}, {Buitrago}  \& {Mortlock}}{{Gr{\"u}tzbauch}
  et~al.}{2011}]{Grutzbauch11}
{Gr{\"u}tzbauch} R.,  {Chuter} R.~W.,  {Conselice} C.~J.,  {Bauer} A.~E.,
  {Bluck} A. F.~L.,  {Buitrago} F.,   {Mortlock} A.,  2011, \mn@doi [\mnras]
  {10.1111/j.1365-2966.2010.18060.x}, \href
  {https://ui.adsabs.harvard.edu/abs/2011MNRAS.412.2361G} {412, 2361}

\bibitem[\protect\citeauthoryear{{Gullieuszik} et~al.,}{{Gullieuszik}
  et~al.}{2015}]{Gullieuszik15}
{Gullieuszik} M.,  et~al., 2015, \mn@doi [\aap] {10.1051/0004-6361/201526061},
  \href {https://ui.adsabs.harvard.edu/abs/2015A&A...581A..41G} {581, A41}

\bibitem[\protect\citeauthoryear{{Gunn} \& {Gott}}{{Gunn} \&
  {Gott}}{1972}]{GG:1972}
{Gunn} J.~E.,  {Gott} J.~R.~I.,  1972, \apj, \href
  {http://adsabs.harvard.edu/abs/1972ApJ...176....1G} {176, 1}

\bibitem[\protect\citeauthoryear{{Guzzo} et~al.,}{{Guzzo}
  et~al.}{2014}]{Guzzo14}
{Guzzo} L.,  et~al., 2014, \mn@doi [\aap] {10.1051/0004-6361/201321489}, \href
  {http://adsabs.harvard.edu/abs/2014A%26A...566A.108G} {566, A108}

\bibitem[\protect\citeauthoryear{{Hahn} et~al.,}{{Hahn}
  et~al.}{2019}]{Hahn2019}
{Hahn} C.,  et~al., 2019, \mn@doi [\apj] {10.3847/1538-4357/aafedd}, \href
  {https://ui.adsabs.harvard.edu/abs/2019ApJ...872..160H} {872, 160}

\bibitem[\protect\citeauthoryear{{Hasinger}}{{Hasinger}}{2008}]{Hasinger:2008}
{Hasinger} G.,  2008, \mn@doi [\aap] {10.1051/0004-6361:200809839}, \href
  {https://ui.adsabs.harvard.edu/abs/2008A&A...490..905H} {490, 905}

\bibitem[\protect\citeauthoryear{{Henriques}, {White}, {Thomas}, {Angulo},
  {Guo}, {Lemson}  \& {Wang}}{{Henriques} et~al.}{2017}]{Henriques17}
{Henriques} B. M.~B.,  {White} S. D.~M.,  {Thomas} P.~A.,  {Angulo} R.~E.,
  {Guo} Q.,  {Lemson} G.,   {Wang} W.,  2017, \mn@doi [\mnras]
  {10.1093/mnras/stx1010}, \href
  {https://ui.adsabs.harvard.edu/abs/2017MNRAS.469.2626H} {469, 2626}

\bibitem[\protect\citeauthoryear{{Hogg} et~al.,}{{Hogg} et~al.}{2002}]{Hogg02}
{Hogg} D.~W.,  et~al., 2002, \mn@doi [\aj] {10.1086/341392}, \href
  {https://ui.adsabs.harvard.edu/abs/2002AJ....124..646H} {124, 646}

\bibitem[\protect\citeauthoryear{{Hopkins}, {Kere{\v{s}}}, {O{\~n}orbe},
  {Faucher-Gigu{\`e}re}, {Quataert}, {Murray}  \& {Bullock}}{{Hopkins}
  et~al.}{2014}]{Hopkins14}
{Hopkins} P.~F.,  {Kere{\v{s}}} D.,  {O{\~n}orbe} J.,  {Faucher-Gigu{\`e}re}
  C.-A.,  {Quataert} E.,  {Murray} N.,   {Bullock} J.~S.,  2014, \mn@doi
  [\mnras] {10.1093/mnras/stu1738}, \href
  {https://ui.adsabs.harvard.edu/abs/2014MNRAS.445..581H} {445, 581}

\bibitem[\protect\citeauthoryear{{Hwang}, {Shin}  \& {Song}}{{Hwang}
  et~al.}{2019}]{Hwang2019}
{Hwang} H.~S.,  {Shin} J.,   {Song} H.,  2019, \mn@doi [\mnras]
  {10.1093/mnras/stz2136}, \href
  {https://ui.adsabs.harvard.edu/abs/2019MNRAS.489..339H} {489, 339}

\bibitem[\protect\citeauthoryear{{Jaff{\'e}}, {Poggianti}, {Verheijen},
  {Deshev}  \& {van Gorkom}}{{Jaff{\'e}} et~al.}{2012}]{Jaffe12}
{Jaff{\'e}} Y.~L.,  {Poggianti} B.~M.,  {Verheijen} M. A.~W.,  {Deshev} B.~Z.,
   {van Gorkom} J.~H.,  2012, \mn@doi [\apjl] {10.1088/2041-8205/756/2/L28},
  \href {https://ui.adsabs.harvard.edu/abs/2012ApJ...756L..28J} {756, L28}

\bibitem[\protect\citeauthoryear{{Juneau} et~al.,}{{Juneau}
  et~al.}{2005}]{Juneau05}
{Juneau} S.,  et~al., 2005, \mn@doi [\apjl] {10.1086/427937}, \href
  {https://ui.adsabs.harvard.edu/abs/2005ApJ...619L.135J} {619, L135}

\bibitem[\protect\citeauthoryear{{Kaiser}}{{Kaiser}}{1986}]{Kaiser:86}
{Kaiser} N.,  1986, \mn@doi [\mnras] {10.1093/mnras/222.2.323}, \href
  {https://ui.adsabs.harvard.edu/abs/1986MNRAS.222..323K} {222, 323}

\bibitem[\protect\citeauthoryear{{Klypin}, {Yepes}, {Gottl{\"o}ber}, {Prada}
  \& {He{\ss}}}{{Klypin} et~al.}{2016}]{Klypin2016}
{Klypin} A.,  {Yepes} G.,  {Gottl{\"o}ber} S.,  {Prada} F.,   {He{\ss}} S.,
  2016, \mn@doi [\mnras] {10.1093/mnras/stw248}, \href
  {https://ui.adsabs.harvard.edu/abs/2016MNRAS.457.4340K} {457, 4340}

\bibitem[\protect\citeauthoryear{{Knebe}, {Gill}, {Gibson}, {Lewis}, {Ibata}
  \& {Dopita}}{{Knebe} et~al.}{2004}]{Knebe:2004}
{Knebe} A.,  {Gill} S. P.~D.,  {Gibson} B.~K.,  {Lewis} G.~F.,  {Ibata} R.~A.,
   {Dopita} M.~A.,  2004, \mn@doi [\apj] {10.1086/381306}, \href
  {https://ui.adsabs.harvard.edu/abs/2004ApJ...603....7K} {603, 7}

\bibitem[\protect\citeauthoryear{{Kodama} et~al.,}{{Kodama}
  et~al.}{2004}]{Kodama04}
{Kodama} T.,  et~al., 2004, \mn@doi [\mnras]
  {10.1111/j.1365-2966.2004.07711.x}, \href
  {https://ui.adsabs.harvard.edu/abs/2004MNRAS.350.1005K} {350, 1005}

\bibitem[\protect\citeauthoryear{{Kuchner} et~al.,}{{Kuchner}
  et~al.}{2022}]{Kuchner:2022}
{Kuchner} U.,  et~al., 2022, \mn@doi [\mnras] {10.1093/mnras/stab3419}, \href
  {https://ui.adsabs.harvard.edu/abs/2022MNRAS.510..581K} {510, 581}

\bibitem[\protect\citeauthoryear{{Lagos}, {Cora}  \& {Padilla}}{{Lagos}
  et~al.}{2008}]{Lagos08}
{Lagos} C. D.~P.,  {Cora} S.~A.,   {Padilla} N.~D.,  2008, \mn@doi [\mnras]
  {10.1111/j.1365-2966.2008.13456.x}, \href
  {https://ui.adsabs.harvard.edu/abs/2008MNRAS.388..587L} {388, 587}

\bibitem[\protect\citeauthoryear{{Larson}, {Tinsley}  \& {Caldwell}}{{Larson}
  et~al.}{1980}]{Larson:1980}
{Larson} R.~B.,  {Tinsley} B.~M.,   {Caldwell} C.~N.,  1980, \mn@doi [\apj]
  {10.1086/157917}, \href {http://adsabs.harvard.edu/abs/1980ApJ...237..692L}
  {237, 692}

\bibitem[\protect\citeauthoryear{{Lilly}, {Le Fevre}, {Hammer}  \&
  {Crampton}}{{Lilly} et~al.}{1996}]{Lilly96}
{Lilly} S.~J.,  {Le Fevre} O.,  {Hammer} F.,   {Crampton} D.,  1996, \mn@doi
  [\apjl] {10.1086/309975}, \href
  {https://ui.adsabs.harvard.edu/abs/1996ApJ...460L...1L} {460, L1}

\bibitem[\protect\citeauthoryear{{Madau} \& {Dickinson}}{{Madau} \&
  {Dickinson}}{2014}]{Madau14}
{Madau} P.,  {Dickinson} M.,  2014, \mn@doi [\araa]
  {10.1146/annurev-astro-081811-125615}, \href
  {https://ui.adsabs.harvard.edu/abs/2014ARA&A..52..415M} {52, 415}

\bibitem[\protect\citeauthoryear{{Mart{\'{\i}}nez}, {Coenda}  \&
  {Muriel}}{{Mart{\'{\i}}nez} et~al.}{2008}]{martinez08}
{Mart{\'{\i}}nez} H.~J.,  {Coenda} V.,   {Muriel} H.,  2008, \mn@doi [\mnras]
  {10.1111/j.1365-2966.2008.13929.x}, \href
  {http://adsabs.harvard.edu/abs/2008MNRAS.391..585M} {391, 585}

\bibitem[\protect\citeauthoryear{{Mart{\'{\i}}nez}, {Muriel}  \&
  {Coenda}}{{Mart{\'{\i}}nez} et~al.}{2016}]{Martinez16}
{Mart{\'{\i}}nez} H.~J.,  {Muriel} H.,   {Coenda} V.,  2016, \mn@doi [MNRAS]
  {10.1093/mnras/stv2295}, \href
  {http://adsabs.harvard.edu/abs/2016MNRAS.455..127M} {455, 127}

\bibitem[\protect\citeauthoryear{{McCarthy}, {Frenk}, {Font}, {Lacey}, {Bower},
  {Mitchell}, {Balogh}  \& {Theuns}}{{McCarthy} et~al.}{2008}]{McCarthy:2008}
{McCarthy} I.~G.,  {Frenk} C.~S.,  {Font} A.~S.,  {Lacey} C.~G.,  {Bower}
  R.~G.,  {Mitchell} N.~L.,  {Balogh} M.~L.,   {Theuns} T.,  2008, \mn@doi
  [\mnras] {10.1111/j.1365-2966.2007.12577.x}, \href
  {https://ui.adsabs.harvard.edu/abs/2008MNRAS.383..593M} {383, 593}

\bibitem[\protect\citeauthoryear{{Merritt}}{{Merritt}}{1983}]{Merritt83}
{Merritt} D.,  1983, \mn@doi [\apj] {10.1086/160571}, \href
  {https://ui.adsabs.harvard.edu/abs/1983ApJ...264...24M} {264, 24}

\bibitem[\protect\citeauthoryear{{Moore}, {Katz}, {Lake}, {Dressler}  \&
  {Oemler}}{{Moore} et~al.}{1996}]{Moore:1996}
{Moore} B.,  {Katz} N.,  {Lake} G.,  {Dressler} A.,   {Oemler} A.,  1996,
  \mn@doi [\nat] {10.1038/379613a0}, \href
  {http://adsabs.harvard.edu/abs/1996Natur.379..613M} {379, 613}

\bibitem[\protect\citeauthoryear{{Moore}, {Lake}, {Quinn}  \& {Stadel}}{{Moore}
  et~al.}{1999}]{Moore:1999}
{Moore} B.,  {Lake} G.,  {Quinn} T.,   {Stadel} J.,  1999, \mn@doi [\mnras]
  {10.1046/j.1365-8711.1999.02345.x}, \href
  {http://adsabs.harvard.edu/abs/1999MNRAS.304..465M} {304, 465}

\bibitem[\protect\citeauthoryear{{Moretti} et~al.,}{{Moretti}
  et~al.}{2017}]{Moretti17}
{Moretti} A.,  et~al., 2017, \mn@doi [\aap] {10.1051/0004-6361/201630030},
  \href {https://ui.adsabs.harvard.edu/abs/2017A&A...599A..81M} {599, A81}

\bibitem[\protect\citeauthoryear{{Mu{\~n}oz Arancibia}, {Navarrete}, {Padilla},
  {Cora}, {Gawiser}, {Kurczynski}  \& {Ruiz}}{{Mu{\~n}oz Arancibia}
  et~al.}{2015}]{MunozArancibia15}
{Mu{\~n}oz Arancibia} A.~M.,  {Navarrete} F.~P.,  {Padilla} N.~D.,  {Cora}
  S.~A.,  {Gawiser} E.,  {Kurczynski} P.,   {Ruiz} A.~N.,  2015, \mn@doi
  [\mnras] {10.1093/mnras/stu2237}, \href
  {https://ui.adsabs.harvard.edu/abs/2015MNRAS.446.2291M} {446, 2291}

\bibitem[\protect\citeauthoryear{{Orsi}, {Padilla}, {Groves}, {Cora}, {Tecce},
  {Gargiulo}  \& {Ruiz}}{{Orsi} et~al.}{2014}]{Orsi14}
{Orsi} {\'A}.,  {Padilla} N.,  {Groves} B.,  {Cora} S.,  {Tecce} T.,
  {Gargiulo} I.,   {Ruiz} A.,  2014, \mn@doi [\mnras] {10.1093/mnras/stu1203},
  \href {https://ui.adsabs.harvard.edu/abs/2014MNRAS.443..799O} {443, 799}

\bibitem[\protect\citeauthoryear{{Paccagnella} et~al.,}{{Paccagnella}
  et~al.}{2016}]{Paccagnella16}
{Paccagnella} A.,  et~al., 2016, \mn@doi [\apjl] {10.3847/2041-8205/816/2/L25},
  \href {https://ui.adsabs.harvard.edu/abs/2016ApJ...816L..25P} {816, L25}

\bibitem[\protect\citeauthoryear{{Paccagnella} et~al.,}{{Paccagnella}
  et~al.}{2017}]{Paccagnella17}
{Paccagnella} A.,  et~al., 2017, \mn@doi [\apj] {10.3847/1538-4357/aa64d7},
  \href {https://ui.adsabs.harvard.edu/abs/2017ApJ...838..148P} {838, 148}

\bibitem[\protect\citeauthoryear{{Paccagnella}, {Vulcani}, {Poggianti},
  {Moretti}, {Fritz}, {Gullieuszik}  \& {Fasano}}{{Paccagnella}
  et~al.}{2019}]{Paccagnella19}
{Paccagnella} A.,  {Vulcani} B.,  {Poggianti} B.~M.,  {Moretti} A.,  {Fritz}
  J.,  {Gullieuszik} M.,   {Fasano} G.,  2019, \mn@doi [\mnras]
  {10.1093/mnras/sty2728}, \href
  {https://ui.adsabs.harvard.edu/abs/2019MNRAS.482..881P} {482, 881}

\bibitem[\protect\citeauthoryear{{Pallero}, {G{\'o}mez}, {Padilla},
  {Torres-Flores}, {Demarco}, {Cerulo}  \& {Olave-Rojas}}{{Pallero}
  et~al.}{2019}]{pallero_2019}
{Pallero} D.,  {G{\'o}mez} F.~A.,  {Padilla} N.~D.,  {Torres-Flores} S.,
  {Demarco} R.,  {Cerulo} P.,   {Olave-Rojas} D.,  2019, \mn@doi [\mnras]
  {10.1093/mnras/stz1745}, \href
  {https://ui.adsabs.harvard.edu/abs/2019MNRAS.488..847P} {488, 847}

\bibitem[\protect\citeauthoryear{{Patel}, {Holden}, {Kelson}, {Illingworth}  \&
  {Franx}}{{Patel} et~al.}{2009}]{Patel09}
{Patel} S.~G.,  {Holden} B.~P.,  {Kelson} D.~D.,  {Illingworth} G.~D.,
  {Franx} M.,  2009, \mn@doi [\apjl] {10.1088/0004-637X/705/1/L67}, \href
  {https://ui.adsabs.harvard.edu/abs/2009ApJ...705L..67P} {705, L67}

\bibitem[\protect\citeauthoryear{{Pearson} et~al.,}{{Pearson}
  et~al.}{2018}]{Pearson18}
{Pearson} W.~J.,  et~al., 2018, \mn@doi [\aap] {10.1051/0004-6361/201832821},
  \href {https://ui.adsabs.harvard.edu/abs/2018A&A...615A.146P} {615, A146}

\bibitem[\protect\citeauthoryear{{Peng} et~al.,}{{Peng} et~al.}{2010}]{Peng10}
{Peng} Y.-j.,  et~al., 2010, \mn@doi [\apj] {10.1088/0004-637X/721/1/193},
  \href {http://adsabs.harvard.edu/abs/2010ApJ...721..193P} {721, 193}

\bibitem[\protect\citeauthoryear{{Pereyra}, {Sgr{\'o}}, {Merch{\'a}n},
  {Stasyszyn}  \& {Paz}}{{Pereyra} et~al.}{2020}]{Pereyra20}
{Pereyra} L.~A.,  {Sgr{\'o}} M.~A.,  {Merch{\'a}n} M.~E.,  {Stasyszyn} F.~A.,
  {Paz} D.~J.,  2020, \mn@doi [\mnras] {10.1093/mnras/staa3112}, \href
  {https://ui.adsabs.harvard.edu/abs/2020MNRAS.499.4876P} {499, 4876}

\bibitem[\protect\citeauthoryear{{Pillepich} et~al.,}{{Pillepich}
  et~al.}{2018}]{illustristng}
{Pillepich} A.,  et~al., 2018, \mn@doi [\mnras] {10.1093/mnras/stx3112}, \href
  {https://ui.adsabs.harvard.edu/abs/2018MNRAS.475..648P} {475, 648}

\bibitem[\protect\citeauthoryear{{Planck Collaboration} et~al.,}{{Planck
  Collaboration} et~al.}{2014}]{Planck14}
{Planck Collaboration} et~al., 2014, \mn@doi [\aap]
  {10.1051/0004-6361/201321591}, \href
  {https://ui.adsabs.harvard.edu/abs/2014A&A...571A..16P} {571, A16}

\bibitem[\protect\citeauthoryear{{Poggianti}, {Smail}, {Dressler}, {Couch},
  {Barger}, {Butcher}, {Ellis}  \& {Oemler}}{{Poggianti}
  et~al.}{1999}]{Poggianti99}
{Poggianti} B.~M.,  {Smail} I.,  {Dressler} A.,  {Couch} W.~J.,  {Barger}
  A.~J.,  {Butcher} H.,  {Ellis} R.~S.,   {Oemler} Augustus J.,  1999, \mn@doi
  [\apj] {10.1086/307322}, \href
  {https://ui.adsabs.harvard.edu/abs/1999ApJ...518..576P} {518, 576}

\bibitem[\protect\citeauthoryear{{Popesso} et~al.,}{{Popesso}
  et~al.}{2011}]{Popesso11}
{Popesso} P.,  et~al., 2011, \mn@doi [\aap] {10.1051/0004-6361/201015672},
  \href {https://ui.adsabs.harvard.edu/abs/2011A&A...532A.145P} {532, A145}

\bibitem[\protect\citeauthoryear{{Rasmussen}, {Ponman}  \&
  {Mulchaey}}{{Rasmussen} et~al.}{2006}]{Rasmussen06}
{Rasmussen} J.,  {Ponman} T.~J.,   {Mulchaey} J.~S.,  2006, \mn@doi [\mnras]
  {10.1111/j.1365-2966.2006.10492.x}, \href
  {https://ui.adsabs.harvard.edu/abs/2006MNRAS.370..453R} {370, 453}

\bibitem[\protect\citeauthoryear{{Riebe} et~al.,}{{Riebe}
  et~al.}{2013}]{Riebe13}
{Riebe} K.,  et~al., 2013, \mn@doi [Astronomische Nachrichten]
  {10.1002/asna.201211900}, \href
  {https://ui.adsabs.harvard.edu/abs/2013AN....334..691R} {334, 691}

\bibitem[\protect\citeauthoryear{{Rost}, {Stasyszyn}, {Pereyra}  \&
  {Mart{\'\i}nez}}{{Rost} et~al.}{2019}]{Rost:2019}
{Rost} A.,  {Stasyszyn} F.,  {Pereyra} L.,   {Mart{\'\i}nez} H.~J.,  2019,
  MNRAS submitted, \href
  {https://ui.adsabs.harvard.edu/abs/2019arXiv191108545R} {p. arXiv:1911.08545}

\bibitem[\protect\citeauthoryear{{Rost}, {Stasyszyn}, {Pereyra}  \&
  {Mart{\'\i}nez}}{{Rost} et~al.}{2020}]{Rost20}
{Rost} A.,  {Stasyszyn} F.,  {Pereyra} L.,   {Mart{\'\i}nez} H.~J.,  2020,
  \mn@doi [\mnras] {10.1093/mnras/staa320}, \href
  {https://ui.adsabs.harvard.edu/abs/2020MNRAS.493.1936R} {493, 1936}

\bibitem[\protect\citeauthoryear{{Salerno}, {Mart{\'\i}nez}  \&
  {Muriel}}{{Salerno} et~al.}{2019}]{Salerno2019}
{Salerno} J.~M.,  {Mart{\'\i}nez} H.~J.,   {Muriel} H.,  2019, \mn@doi [\mnras]
  {10.1093/mnras/sty3456}, \href
  {https://ui.adsabs.harvard.edu/abs/2019MNRAS.484....2S} {484, 2}

\bibitem[\protect\citeauthoryear{{Salerno} et~al.,}{{Salerno}
  et~al.}{2020}]{Salerno2020}
{Salerno} J.~M.,  et~al., 2020, \mn@doi [\mnras] {10.1093/mnras/staa545}, \href
  {https://ui.adsabs.harvard.edu/abs/2020MNRAS.493.4950S} {493, 4950}

\bibitem[\protect\citeauthoryear{{Sarron}, {Adami}, {Durret}  \&
  {Laigle}}{{Sarron} et~al.}{2019}]{Sarron:2019}
{Sarron} F.,  {Adami} C.,  {Durret} F.,   {Laigle} C.,  2019, \mn@doi [\aap]
  {10.1051/0004-6361/201935394}, \href
  {https://ui.adsabs.harvard.edu/abs/2019A&A...632A..49S} {632, A49}

\bibitem[\protect\citeauthoryear{{Schaye} et~al.,}{{Schaye}
  et~al.}{2015}]{eagle_1}
{Schaye} J.,  et~al., 2015, \mn@doi [\mnras] {10.1093/mnras/stu2058}, \href
  {https://ui.adsabs.harvard.edu/abs/2015MNRAS.446..521S} {446, 521}

\bibitem[\protect\citeauthoryear{{Scodeggio} et~al.,}{{Scodeggio}
  et~al.}{2018}]{vipers}
{Scodeggio} M.,  et~al., 2018, \mn@doi [\aap] {10.1051/0004-6361/201630114},
  \href {https://ui.adsabs.harvard.edu/abs/2018A&A...609A..84S} {609, A84}

\bibitem[\protect\citeauthoryear{{Silverman} et~al.,}{{Silverman}
  et~al.}{2008}]{Silverman:2008}
{Silverman} J.~D.,  et~al., 2008, \mn@doi [\apj] {10.1086/527283}, \href
  {https://ui.adsabs.harvard.edu/abs/2008ApJ...675.1025S} {675, 1025}

\bibitem[\protect\citeauthoryear{{Smith} et~al.,}{{Smith}
  et~al.}{2015}]{Smith:2015}
{Smith} R.,  et~al., 2015, \mn@doi [\mnras] {10.1093/mnras/stv2082}, \href
  {https://ui.adsabs.harvard.edu/abs/2015MNRAS.454.2502S} {454, 2502}

\bibitem[\protect\citeauthoryear{{Springel}, {Yoshida}  \& {White}}{{Springel}
  et~al.}{2001}]{Springel01}
{Springel} V.,  {Yoshida} N.,   {White} S. D.~M.,  2001, \mn@doi [\na]
  {10.1016/S1384-1076(01)00042-2}, \href
  {https://ui.adsabs.harvard.edu/abs/2001NewA....6...79S} {6, 79}

\bibitem[\protect\citeauthoryear{{Stevens} \& {Brown}}{{Stevens} \&
  {Brown}}{2017}]{Stevens17}
{Stevens} A. R.~H.,  {Brown} T.,  2017, \mn@doi [\mnras]
  {10.1093/mnras/stx1596}, \href
  {https://ui.adsabs.harvard.edu/abs/2017MNRAS.471..447S} {471, 447}

\bibitem[\protect\citeauthoryear{{Strateva} et~al.,}{{Strateva}
  et~al.}{2001}]{Strateva01}
{Strateva} I.,  et~al., 2001, \mn@doi [\aj] {10.1086/323301}, \href
  {https://ui.adsabs.harvard.edu/abs/2001AJ....122.1861S} {122, 1861}

\bibitem[\protect\citeauthoryear{{Stringer}, {Bower}, {Cole}, {Frenk}  \&
  {Theuns}}{{Stringer} et~al.}{2012}]{Stringer:2012}
{Stringer} M.~J.,  {Bower} R.~G.,  {Cole} S.,  {Frenk} C.~S.,   {Theuns} T.,
  2012, \mn@doi [\mnras] {10.1111/j.1365-2966.2012.20982.x}, \href
  {https://ui.adsabs.harvard.edu/abs/2012MNRAS.423.1596S} {423, 1596}

\bibitem[\protect\citeauthoryear{{Taylor} et~al.,}{{Taylor}
  et~al.}{2015}]{Taylor15}
{Taylor} E.~N.,  et~al., 2015, \mn@doi [\mnras] {10.1093/mnras/stu1900}, \href
  {https://ui.adsabs.harvard.edu/abs/2015MNRAS.446.2144T} {446, 2144}

\bibitem[\protect\citeauthoryear{{Tecce}, {Cora}, {Tissera}, {Abadi}  \&
  {Lagos}}{{Tecce} et~al.}{2010}]{Tecce10}
{Tecce} T.~E.,  {Cora} S.~A.,  {Tissera} P.~B.,  {Abadi} M.~G.,   {Lagos} C.
  D.~P.,  2010, \mn@doi [\mnras] {10.1111/j.1365-2966.2010.17262.x}, \href
  {https://ui.adsabs.harvard.edu/abs/2010MNRAS.408.2008T} {408, 2008}

\bibitem[\protect\citeauthoryear{{Trujillo-Gomez}, {Klypin}, {Col{\'\i}n},
  {Ceverino}, {Arraki}  \& {Primack}}{{Trujillo-Gomez}
  et~al.}{2015}]{Trujillo-Gomez15}
{Trujillo-Gomez} S.,  {Klypin} A.,  {Col{\'\i}n} P.,  {Ceverino} D.,  {Arraki}
  K.~S.,   {Primack} J.,  2015, \mn@doi [\mnras] {10.1093/mnras/stu2037}, \href
  {https://ui.adsabs.harvard.edu/abs/2015MNRAS.446.1140T} {446, 1140}

\bibitem[\protect\citeauthoryear{{Vega-Mart{\'\i}nez}, {G{\'o}mez}, {Cora}  \&
  {Hough}}{{Vega-Mart{\'\i}nez} et~al.}{2022}]{VegaMartinez22}
{Vega-Mart{\'\i}nez} C.~A.,  {G{\'o}mez} F.~A.,  {Cora} S.~A.,   {Hough} T.,
  2022, \mn@doi [\mnras] {10.1093/mnras/stab2908}, \href
  {https://ui.adsabs.harvard.edu/abs/2022MNRAS.509..701V} {509, 701}

\bibitem[\protect\citeauthoryear{{Vijayaraghavan} \& {Ricker}}{{Vijayaraghavan}
  \& {Ricker}}{2015}]{Vijayaraghavan:2015}
{Vijayaraghavan} R.,  {Ricker} P.~M.,  2015, \mn@doi [\mnras]
  {10.1093/mnras/stv476}, \href
  {https://ui.adsabs.harvard.edu/abs/2015MNRAS.449.2312V} {449, 2312}

\bibitem[\protect\citeauthoryear{{Villalobos}, {De Lucia}  \&
  {Murante}}{{Villalobos} et~al.}{2014}]{Villalobos:2014}
{Villalobos} {\'A}.,  {De Lucia} G.,   {Murante} G.,  2014, \mn@doi [\mnras]
  {10.1093/mnras/stu1278}, \href
  {https://ui.adsabs.harvard.edu/abs/2014MNRAS.444..313V} {444, 313}

\bibitem[\protect\citeauthoryear{{Vulcani}, {Poggianti}, {Finn}, {Rudnick},
  {Desai}  \& {Bamford}}{{Vulcani} et~al.}{2010}]{Vulcani10}
{Vulcani} B.,  {Poggianti} B.~M.,  {Finn} R.~A.,  {Rudnick} G.,  {Desai} V.,
  {Bamford} S.,  2010, \mn@doi [\apjl] {10.1088/2041-8205/710/1/L1}, \href
  {https://ui.adsabs.harvard.edu/abs/2010ApJ...710L...1V} {710, L1}

\bibitem[\protect\citeauthoryear{{Walters}, {Woo}  \& {Ellison}}{{Walters}
  et~al.}{2022}]{walters_2022}
{Walters} D.,  {Woo} J.,   {Ellison} S.~L.,  2022, \mn@doi [\mnras]
  {10.1093/mnras/stac283}, \href
  {https://ui.adsabs.harvard.edu/abs/2022MNRAS.tmp..293W} {}

\bibitem[\protect\citeauthoryear{{Wang} et~al.,}{{Wang}
  et~al.}{2018}]{wang_2018}
{Wang} Y.,  et~al., 2018, \mn@doi [\apj] {10.3847/1538-4357/aae52e}, \href
  {https://ui.adsabs.harvard.edu/abs/2018ApJ...868..130W} {868, 130}

\bibitem[\protect\citeauthoryear{{Wetzel}, {Tinker}  \& {Conroy}}{{Wetzel}
  et~al.}{2012}]{Wetzel1012}
{Wetzel} A.~R.,  {Tinker} J.~L.,   {Conroy} C.,  2012, \mn@doi [\mnras]
  {10.1111/j.1365-2966.2012.21188.x}, \href
  {https://ui.adsabs.harvard.edu/abs/2012MNRAS.424..232W} {424, 232}

\bibitem[\protect\citeauthoryear{{Xie}, {De Lucia}, {Hirschmann}  \&
  {Fontanot}}{{Xie} et~al.}{2020}]{Xie20}
{Xie} L.,  {De Lucia} G.,  {Hirschmann} M.,   {Fontanot} F.,  2020, \mn@doi
  [\mnras] {10.1093/mnras/staa2370}, \href
  {https://ui.adsabs.harvard.edu/abs/2020MNRAS.498.4327X} {498, 4327}

\bibitem[\protect\citeauthoryear{{York} et~al.,}{{York}
  et~al.}{2000}]{York:2000}
{York} D.~G.,  et~al., 2000, \mn@doi [\aj] {10.1086/301513}, \href
  {https://ui.adsabs.harvard.edu/abs/2000AJ....120.1579Y} {120, 1579}

\makeatother
\end{thebibliography}
%%%%%%%%%%%%%%%%%%%%%%%%%%%%%%%%%%%%%%%%%%%%%%%%%%

%%%%%%%%%%%%%%%%% APPENDICES %%%%%%%%%%%%%%%%%%%%%

\appendix

%%%%%%%%%%%%%%%%%%%%%%%%%%%%%%%%%%%%%%%%%%%%%%%%%%

% Don't change these lines
\bsp	% typesetting comment
\label{lastpage}
	\end{document}